\documentclass[letterpaper,twocolumn,10pt]{article}
\usepackage[]{usenix2021_SOUPS}
\usepackage{float}

\usepackage{breakurl}
\hypersetup{breaklinks=true}

\usepackage{glossaries}
\usepackage{soul}
\usepackage{colortbl}

\def\BibTeX{{\rm B\kern-.05em{\sc i\kern-.025em b}\kern-.08em
        T\kern-.1667em\lower.7ex\hbox{E}\kern-.125emX}}

 \usepackage{titlesec}
 \titleformat{\section}[hang]{\bfseries\sffamily\large}{\thesection}{2ex}{}[]
 \titleformat{\subsection}[hang]{\normalfont\sffamily\fontsize{11}{12.5}\bfseries}{\thesubsection}{2ex}{}[]

\usepackage{subcaption}

\usepackage{csquotes}

\usepackage{pifont}
\usepackage{booktabs,array}
\usepackage{multirow}
\usepackage{pgfplotstable}
\usepackage{multicol}

\usepackage{appendix}

\usepackage[inline]{enumitem}

\usepackage{pdfpages}

\usepackage{graphicx,calc}
\newlength\myheight
\newlength\mydepth
\settototalheight\myheight{Xygp}
\newcommand*\inlinegraphics[1]{%
  \settototalheight\myheight{Xygp}%
  \settodepth\mydepth{Xygp}%
  \raisebox{-0.5\mydepth}{\includegraphics[height=\myheight]{#1}}%
}
\newcommand{\apple}{\inlinegraphics{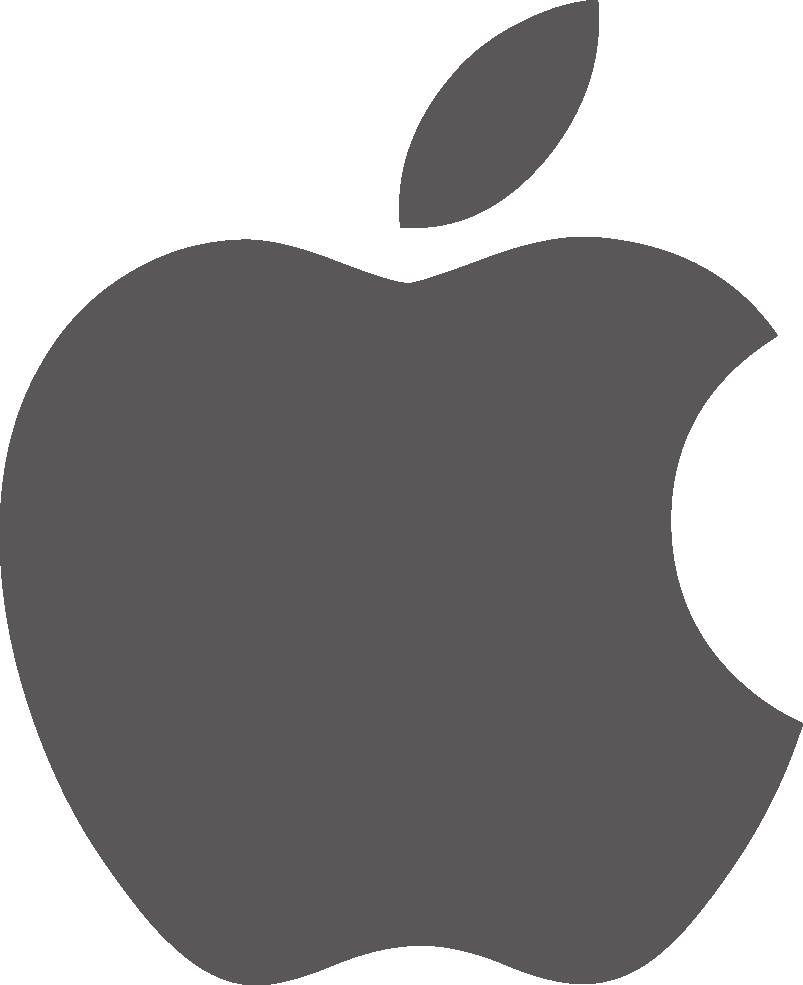}}
\newcommand{\android}{\inlinegraphics{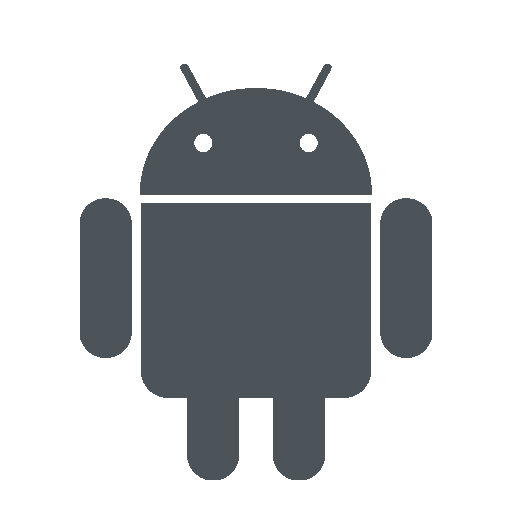}}

\usepackage{comment}

\pgfplotsset{compat=1.8}
\definecolor{rulecolor}{RGB}{0,71,171}
\definecolor{tableheadcolor}{gray}{0.92}
\newcommand{\cmark}{\ding{51}}%
\newcommand{\xmark}{\ding{55}}%

\newcommand{\topline}{ %
        \arrayrulecolor{rulecolor}\specialrule{0.1em}{\abovetopsep}{0pt}%
        \arrayrulecolor{tableheadcolor}\specialrule{\belowrulesep}{0pt}{0pt}%
        \arrayrulecolor{rulecolor}}
\newcommand{\midtopline}{ %
        \arrayrulecolor{tableheadcolor}\specialrule{\aboverulesep}{0pt}{0pt}
        \arrayrulecolor{rulecolor}\specialrule{\lightrulewidth}{0pt}{0pt}%
        \arrayrulecolor{white}\specialrule{\belowrulesep}{0pt}{0pt}%
        \arrayrulecolor{rulecolor}}
\newcommand\topmidheader[2]{\multicolumn{#1}{c}{\textsc{#2}}\\%
                \addlinespace[0.5ex]}

\renewcommand{\toprule}{\topline}

\newcommand{\minihead}[1]{\textbf{#1.}}
 
\usepackage{hhline}
\usepackage{pgf}
\usepackage{collcell}
 
\usepackage{pgfplots}
\DeclareUnicodeCharacter{2212}{−}
\usepgfplotslibrary{groupplots,dateplot}
\usetikzlibrary{patterns,shapes.arrows}
\pgfplotsset{compat=newest}

\usepackage{enumitem}
 
\usepackage[nameinlink,capitalise]{cleveref}

\crefname{lstlisting}{listing}{listings}
\Crefname{lstlisting}{Listing}{Listings}
\crefalias{stepenumi}{step}
\crefname{figure}{Figure}{Figures}
\crefname{equation}{Equation}{Equations}
\crefname{step}{Step}{Steps}
\crefname{appsec}{Appendix}{Appendices}
\crefname{line}{Line}{Lines}

\definecolor{raven}{RGB}{55,61,63}
\definecolor{linkcolor}{rgb}{0.65,0,0}
\definecolor{citecolor}{rgb}{0,0,0.5}
\definecolor{urlcolor}{rgb}{0,0,0.65}
\definecolor{Gray}{gray}{0.9}

\newacronym{vdp}{VDP}{vulnerability disclosure program} 
\newacronym{psirt}{PSIRT}{Product Security Incident Response Team}
\newacronym{cve}{CVE}{Common Vulnerabilities and Exposures}
\newacronym{nvd}{NVD}{National Vulnerability Database}
\newacronym{cvss}{CVSS}{Common Vulnerability Scoring System}
\newacronym{aslr}{ASLR}{Address Space Layout Randomization}

\usepackage{xpatch}
\makeatletter
\xpatchcmd\@array
  {\expandafter\ifx\d@llarbegin\begingroup\hskip-\col@sep\fi}
  {}
  {}{}
\xpatchcmd\@array
  {\expandafter\ifx\d@llarbegin\begingroup\hskip-\col@sep\fi}
  {}
  {}{}
\makeatother

\renewcommand{\labelitemi}{\textbullet}

\usepackage{xspace}
\newcommand{\ninstructors}[0]{49\xspace}
\newcommand{\nadministrators}[0]{14\xspace}

\begin{document}

\date{}

\title{\Large \bf Virtual Classrooms and Real Harms: Remote Learning at U.S. Universities}
\pagenumbering{gobble}


\def\plainauthor{Shaanan Cohney, Ross Teixeira, Anne Kohlbrenner, Arvind Narayanan, Mihir Kshirsagar, Yan Shvartzshnaider, Madelyn Sanfilippo}

\author{
        {\rm Shaanan Cohney}\\
        Princeton University / University of Melbourne
        \and
        {\rm Ross Teixeira}\\
        Princeton University
        \and
        {\rm Anne Kohlbrenner}\\
        Princeton University
        \and
        {\rm Arvind Narayanan}\\
        Princeton University
        \and
        {\rm Mihir Kshirsagar}\\
        Princeton University
        \and
        {\rm Yan Shvartzshnaider}\\
        Princeton University / York University
        \and
        {\rm Madelyn Sanfilippo}\\
        Princeton University / UIUC
}

\maketitle
\thecopyright

\begin{abstract}
        {Universities have been forced to rely on remote educational technology to facilitate the rapid shift to online learning. In doing so, they acquire new risks of security vulnerabilities and privacy violations. To help universities navigate this landscape, we develop a model that describes the actors, incentives, and risks, informed by surveying 49 instructors and 14 administrators at U.S. universities. Next, we develop a methodology for administrators to assess security and privacy risks of these products. We then conduct a privacy and security analysis of 23 popular platforms using a combination of sociological analyses of privacy policies and 129 state laws, alongside a technical assessment of platform software. Based on our findings, we develop recommendations for universities to mitigate the risks to their stakeholders.}
\end{abstract}



\section{Introduction}
The COVID-19 pandemic pushed universities to adopt remote educational platforms. But most of these platforms were not designed with universities in mind. While these platforms allowed institutions to fill an urgent need, they caused novel and well-publicized security and privacy problems. We examine the underlying causes of these problems through an interdisciplinary lens that identifies the institutional structures that make these incidents more likely, and surfaces the tensions between educational goals and the incentives of the software platforms.

We begin in \cref{sec:model} by documenting how different actors in educational settings---students, instructors, and administrators---each bring their own set of preferences and concerns about platforms. These concerns conflict, leaving instructors and students frustrated that platforms do not meet their needs. We use qualitative surveys of \ninstructors instructors and \nadministrators administrators from U.S. universities to help model the considerations. Next, in \cref{sec:privacy} we discuss the risks that emerge from complex social interactions between the actors in our model through the lens of Contextual Integrity (CI)~\cite{nissenbaum2009privacy}, and discuss where the pandemic has disrupted norms for appropriate information flows. In \cref{sec:security} we discuss the security threats that compromise digital systems through unauthorized access. We discuss related work in \cref{sec:related} and future work in \cref{sec:future}. We synthesize our analyses into recommendations for universities and regulators in \cref{sec:discussion}.

Our analysis identifies three factors that contribute to privacy and security problems. First, there are unresolved tensions between the needs of the different stakeholders. For example, instructors' stated preferences for students' cameras to be left on conflicts with students' privacy concerns about misuse of their video feeds.
Second, there are significant gaps between users' preferences for platform behavior and the actual practices of the platform. A recurring theme in our survey was that defaults matter. While developers may include a configuration toggle for stricter security settings (such as storing call recordings locally vs. to a cloud), if this toggle is not turned on by default or communicated clearly to instructors, its usefulness is significantly hampered. A related issue is that faculty often adopt tools on their own initiative, outside official procurement processes which deprives them of protections afforded to large organizations.
Third, there is a regulatory gap created because the existing educational privacy and data security regulations were written for an era of paper-based records in physical classrooms and are not a good fit for regulating practices arising out of remote learning.

\subsubsection*{Contributions:}

\begin{itemize}
    \item We build a novel threat model that represents the stakeholders in virtual classrooms, their interactions, and privacy/security risks. We ground our model in two qualitative surveys we conduct of instructors and college and IT administrators at U.S. universities.
    \item We assess the privacy and security practices of 23 of the most popular platforms identified in our survey. In particular, we find notable differences between the practices of platforms operated under contract with universities, and the same platforms provided to users free of charge. We observe that contracts negotiated with universities result in significant differences in how data is handled by the platforms. We find that these Data Protection Addenda (DPAs) are a powerful tool to shape platform behavior towards the interests of the stakeholders.
    \item We use our research to provide policy guidance. We recommend that universities prioritize developing tools to incorporate continual improvements based on user feedback and allowing instructors to select features that are relevant to individual educational missions and protecting the interests of vulnerable groups. We also recommend strengthening regulatory mechanisms to provide appropriate baseline privacy and security protections.
\end{itemize}



\section{Actors, Incentives, \& Risks}
\label{sec:model}
In education technology, the actors within the system are both principals to protect and potential threats to mitigate. A useful threat model should therefore specify the level of trust to assign to each actor, in recognition of the variety of roles that individuals in that class may play.


Our construction of a threat model is further complicated by the fact that a platform component may seem both a `feature' and a `threat' to different actors (e.g. video recording). Platform developers must not only build a secure product, but one that mediates between the different interests of their users. As a result, our model moves beyond a trusted/untrusted dichotomy to examine the incentives and interests of all actors.

We begin by describing our threat model to understand how the participating actors view their interests. Next, we model the different risks that platforms must mitigate to help fulfill the educational mission of the software. Finally, we conclude by analyzing the survey results that inform our threat model. Our threat model is based on frameworks such as STRIDE~\cite{STRIDE}, socio-technical system analysis~\cite{lock2010}, and Contextual Integrity~\cite{nissenbaum2009privacy}.

\subsection{Actors} We divide the participating actors into internal actors (students, instructors, administrators), and external actors (third parties and adversaries). While each \textit{class} of internal actors has incentive to maintain the integrity of the educational system, we recognize that all internal actors may engage in adversarial behavior. In addition, our survey shows how the priorities of the internal actors may differ---leading to behavior by one internal actor which may be considered adversarial by another actor.



\noindent\minihead{Incentives} Students enroll for reasons beyond the pursuit of education goals. Further, diplomas have a credentialing function which, for some students, may incentivize cheating. Mixed incentives may therefore cause a student to act contrary to other stakeholders' interests.

Administrators who do not personally use the digital classroom may be more willing to sacrifice the privacy concerns of students and instructors for institutional concerns such as cost efficiencies and auditing or reporting capabilities. Administrators commonly endorse cloud solutions such as Canvas which monetize aggregate data about students, representative of trade-offs between different actors' priorities.

While our model groups instructors and administrators together, they are not homogeneous. Instructors and administrators may play different roles in configuring and using platforms. Those administrators whose roles focus on compliance and risk assessment have incentives to prioritize security and privacy concerns, while others prioritize usability and teaching outcomes. At different institutions, various groups may participate in platform procurement, including institution-level administrators, department staff, and instructors, and power dynamics between these groups may also differ. This also highlights a limitation of our model, which draws boundaries between instructors, administrators, and students, when there are often instances in which those roles overlap---for example, graduate students who teach or TA or senior faculty who serve as administrators.

\begin{figure}[t]
    \centering
    \includegraphics[width=.8\columnwidth]{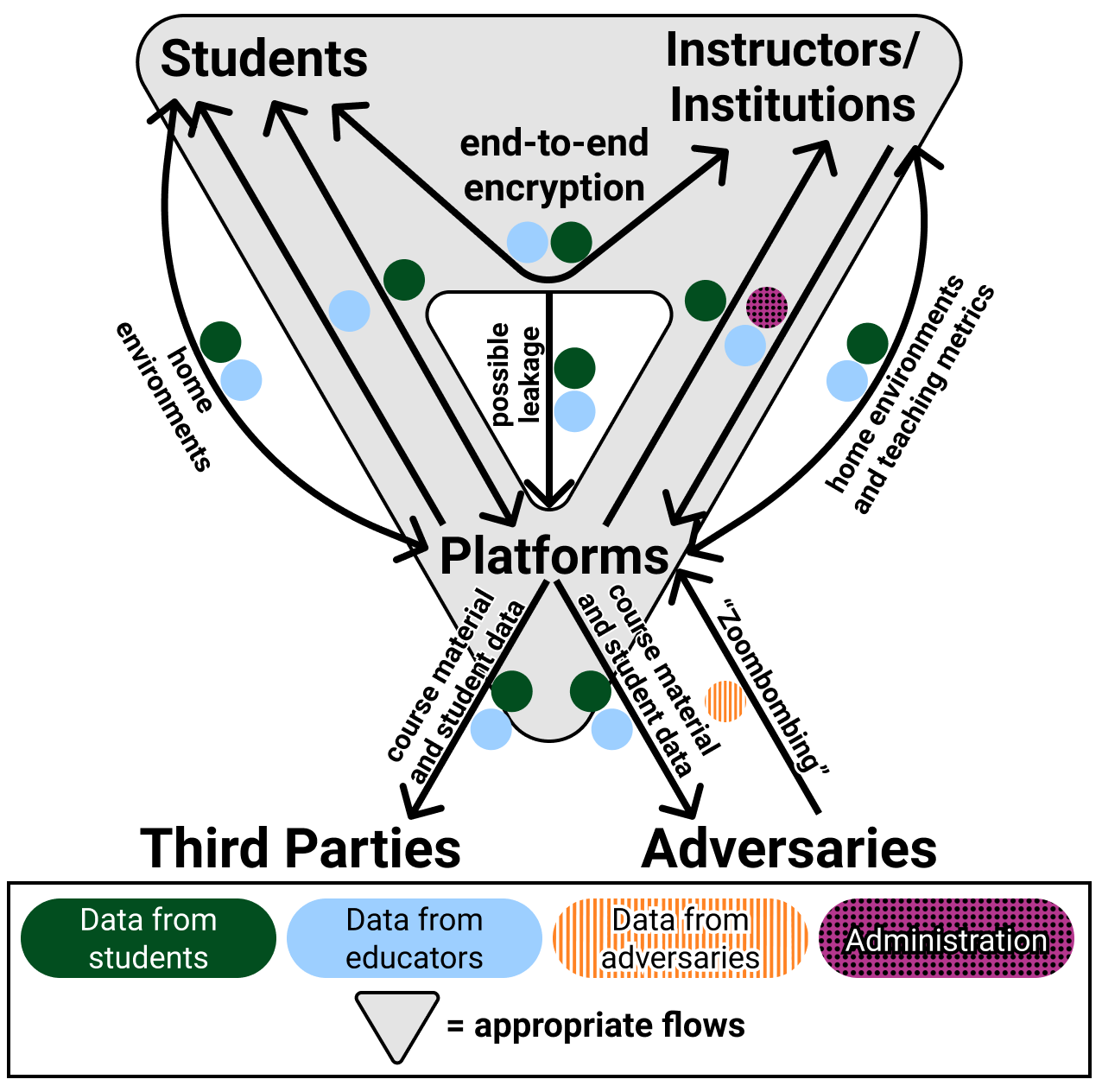}
    \caption{We model typical data flows between stakeholders as mediated by online platforms. A flow is considered appropriate if it corresponds to a legitimate flow under the Contextual Integrity framework we develop in \cref{sec:privacy}. Data from end-to-end encrypted sessions may leak to platform due to poor configuration, implementation, or metadata collection.
    }
    \label{fig:threat-model}
\end{figure}
%
%
 
Under these assumptions, we produce the following set of descriptions about the behaviors of our actors:
\begin{itemize}
   \item \textbf{Students} authenticate their identities and participate in courses by joining live virtual sessions and submitting work. Students may also collaborate and share work with each other. Students may share extraneous information---including their home environment---through video conferences that propagate to instructors (and which may leak to platforms if conferences are not end-to-end encrypted). We do not assume that students are trusted: actors with student-level permissions may try to access or change data not authorized for them, or to prevent access to systems. We also note that students may participate in privacy violations~\cite{ling2020first}. Violations may be intentional when students exploit data-rich platforms or unintentional when platforms leak sensitive information (such as indicators of socio-economic status leaked via video streams).
   \item \textbf{Instructors and administrative staff} generally manage instances of the various platforms (such as individual chat rooms/streams), and are thus significantly privileged and trusted. Well-intentioned instructors may inadvertently breach student or institutional expectations of privacy in a virtual classroom, while misinformed administrators may improperly use student data or metadata to harm students~\cite{citron2021privacy}, such as through false accusations of cheating~\cite{nytimes-dartmouth}. Similarly to students, instructors can precipitate security breaches if their accounts are over privileged, and may prefer to keep extraneous information from being shared with students and administrators (including their own home environments and teaching metrics).
   \item \textbf{Service providers and their third-party affiliates} act to maximize their economic interest within the bounds of their contractual and legal obligations. Platforms may share metadata with third parties for advertising and other business purposes, and course material for services like captioning.
   \item \textbf{External adversaries} may seek to steal student data and course content, and may interfere with live classes (``Zoombombing''). Adversaries may act for profit, entertainment, or other motives.
\end{itemize}
\cref{fig:threat-model} depicts these actors and their interactions.


\subsection{Survey}
\label{sec:model:threats}
We built our threat model using a survey of \ninstructors instructors at U.S. universities to learn what remote learning platforms they use, the features they value in a platform, and their concerns (and those of their students), with particular emphasis on privacy and security. We recruited participants through a public Slack group for instructors teaching remotely, as well a public social media post. 

Separately, we surveyed \nadministrators U.S. university administrators about their schools' procurement processes for new learning platforms. As administrators with influence and understanding of the procurement process are harder to reach, our sample size was limited to \nadministrators participants.

Our institutional review board (IRB) reviewed and approved our study design, consent, and recruitment procedures. All participants affirmatively consented to participate in the study, after reviewing a form approved by the IRB.

The surveys provide a qualitative framing for our platform analyses and recommendations later in the paper. Additional details regarding survey materials and responses are provided in \cref{app:survey:breakdown}.

\begin{figure*}[t]
    \includegraphics[width=.9\textwidth]{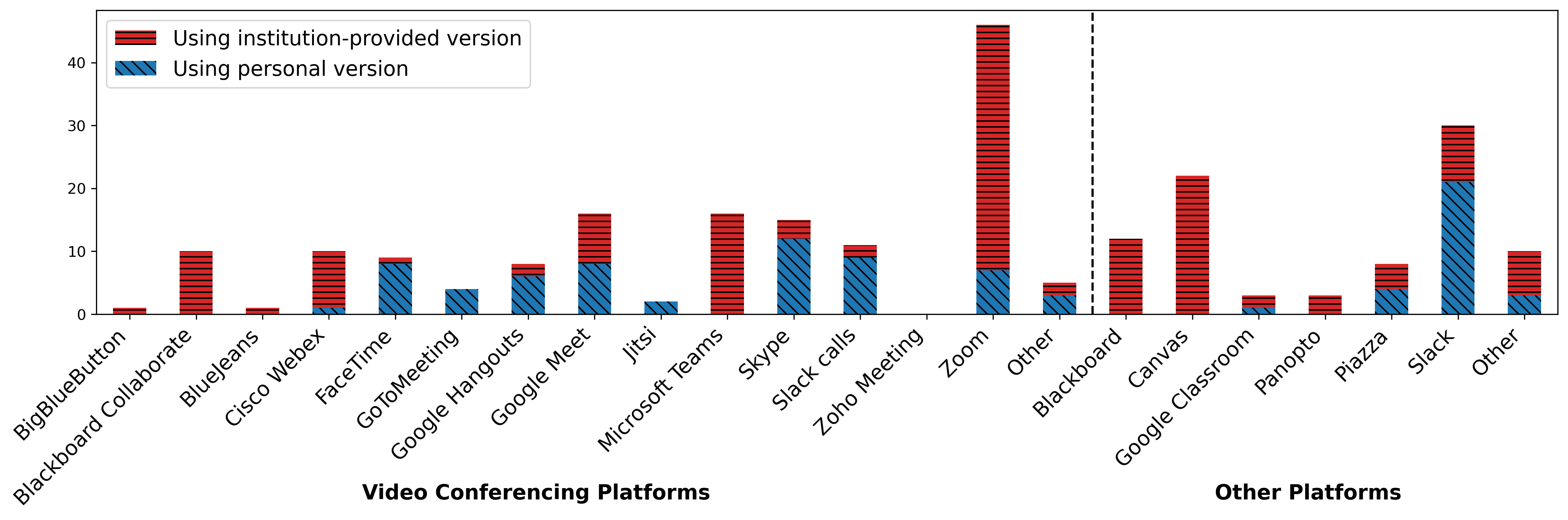}
    \caption{Survey usage counts for remote teaching platforms, including institution-supported versions and personal versions. 
    }
    \label{fig:platform-versions}
\end{figure*}

\subsubsection{Instructor Survey Results}
We report the number of instructors using each platform, as well as whether they use a personal version or an institutionally provided version in \cref{fig:platform-versions}.

We find that the concerns of instructors generally touch on three major themes, with some instructors mentioning multiple themes. First, students' personal data is more easily captured and visible to instructors, platforms, co-inhabitants, and other students from a virtual classroom. This includes students broadcasting their home environments (and socioeconomic indicators therein) through video, private chats which may leak to instructors, and co-inhabitants hearing sensitive class material. It also includes proctoring services that hijack students' computers while monitoring their environment.

Second, personal data is more easily disseminated by platforms to third parties, with little recourse for students or instructors who wish to limit data sharing. instructors were concerned that platforms may sell metadata from students/instructors to advertisers or leak data to services like video captioning, and that law enforcement may request student data from platforms.

Third, platforms are vulnerable to attack from unintended adversaries that threaten to steal data and interfere with courses, due to poor authentication and other security measures by platforms.

Specific concerns and desires mentioned in our instructor survey are listed below.

\textbf{Security/privacy.} 25 (51\%) instructors marked "Yes" to having security and/or privacy concerns with platforms. Respondents did not clearly delineate between security and privacy: notably, each of our five freeform questions, including those unrelated to privacy or security, received at least two responses that we manually coded as pertaining to a privacy or security concern. Fifteen discussed concerns that ``private'' chats or videos were not private, especially to other students; nine were concerned that platforms did not adequately secure meetings against intrusion; and two mentioned concerns about theft of intellectual property.
 
\textbf{Surveillance.} Instructors were concerned about the surveillance implications of using remote learning tools, both for student data and their own. For example, five instructors noted they were concerned platforms might share data with third parties, with particular emphasis on course data and personal data of students. Another instructor also denounced potential data sharing with law enforcement. Meanwhile, two instructors mentioned not wanting to share their own home environment with students, and one was concerned that their campus would use platform metrics to judge their teaching performance. One instructor even tried to use Privacy Badger~\cite{privacy-badger} to disable tracking on Blackboard, but this interfered with classroom functions. 

\textbf{Recording restrictions.} Nine instructors reported it was important that platforms provided the ability to save recordings locally or to private clouds. One advocated for allowing students to ``opt out'' of showing their video on recordings. Together, these respondents discussed restrictions on every aspect of class recordings: restricting who can make recordings, who hosts them, and who is allowed to access them.

\textbf{Platform choice.} Respondents also reported specific dissatisfaction with their institutions in selecting and configuring platforms. Seven instructors reported frustration with the choice of platforms by their institutions, with one instructor noting that Canvas' `testing capabilities are not as nice as' a platform used at another university. Two of these instructors discussed using alternative, free software such as Humanities Commons and Mattermost. The other instructor noted a stored message limit in the free version of Slack, which demonstrates an issue for instructors using software that is not supported by their institutions. One instructor also reported ``We don't have the [Canvas] version with video conferencing,'' whereas conferences are a configurable permission by universities at no extra cost~\cite{canvas-conference}. This highlights that instructors and universities may not be aware of configurable or default settings in platforms.

\subsubsection{Administrator Survey Results}
While our administrator survey had only \nadministrators respondents, it surfaced issues about how their institutions select software---issues other universities are likely to share. Eight administrators reported that their institutions did not have a formal process for selecting new platforms, while one reported a price threshold for invoking a more formal process. Administrators marked that the platform selection process was driven primarily by Faculty and IT staff. Two administrators reported having processes for collecting feedback from faculty on learning platforms, but one emphasized that re-evaluation of whether a platform met needs would only happen at the time of “contract renewal.” Further, “legal obligations” were rated moderately important (when rated from `not at all' to `extremely' important) by 3/4 administrators who answered this question, implying that other actors are likely responsible for compliance. Four administrators rated a platform's privacy features as `very' or `extremely' important. Three of those four rated security equally important, while one rated security only `slightly' important. Three administrators described negotiating for more privacy guarantees such as DUO and HIPAA compliance.

According to one administrator, the most significant change relative to COVID-19 was not procurement of additional video platforms, but filling new needs such as proctoring software for ``high-stakes testing.’’ 

\section{Privacy Analysis}
\label{sec:privacy}
As the cost of data collection from online interactions is low and there is less friction to collect such data compared to the offline context, the platforms end up collecting vast amounts of data. As discussed below, these practices, despite ostensibly being compliant with the existing regulations, can conflict with with contextual educational privacy norms and expectations.

\subsection{Empirical Approach to Privacy Analysis}
\label{sec:privacy:empirical}

Given that the expectations and interests of the relevant actors are complex, conflicting, and overlapping (as discussed relative to incentives in \cref{sec:model:threats}), we adopt descriptive institutional analysis frameworks~\cite{frischmann2014governing,sanfilippo2018commons} to structure our governance inquiries. This approach recognizes and builds upon a conceptualization of technology governance as an assemblage of laws, norms, markets, and architecture~\cite{haynes2016,lessig,freiwald}. We also use the Contextual Integrity framework to understand how stakeholder expectations change when the physical classroom becomes digital, drawing on established methods~\cite{shvartzshnaider2016learning}. CI views privacy as the appropriate flow of information, where appropriateness is defined by the governing contextual norms. We draw on the survey responses to identify which information handling practices are appropriate in the educational context.

We employ the existing integrated GKC-CI codebook~\cite{sanfilippo2020disaster} to operationalize these frameworks and assess governance of information flows associated with the platforms. Two of the investigators assessed inter-rater reliability in two phases, based on a subset of privacy policies. After the first round, overall agreement was 86.27\% with ranges in Krippendorf's alpha from .49 to .97, with 2 of 8 codes not reaching the required .8 threshold. Following inter-rater discussion, a revised second round of coding was conducted wherein agreement improved to 93.67\% overall, with Krippendorf's alpha ranging from .81 to .97, indicating excellent agreement. The same investigators subsequently applied the codebook to information flows and governance described in DPAs and regulations.

\subsection{Law}

We observe that current laws do not sufficiently control platform behavior to conform to the privacy norms of higher education. While the market for educational technology is comparatively highly regulated by state and federal laws, those laws are not always effective. For a university to control a platform's information practices in a way that fits within the spirit (if not the actual application) of federal and state laws, it must take active and intentional supplementary governance interventions, such as by customizing DPA as we discuss in \cref{sec:privacy:empirical}.

\subsubsection{Background}

We provide a brief overview of the primary legal frameworks in the United States that apply specifically to student privacy.

\minihead{Federal Regulation} The Family Educational Rights and Privacy Act (FERPA)~\cite{ferpa} protects defined categories of student records and enrollment at an educational institution~\cite{ferpaoverview}. The law, enacted in 1974, was designed for a paper-based record system with discrete and limited set of records. 

FERPA requires schools and universities to keep records of each external disclosure of student information and requires the records to be available on request by the subject~\cite{ferpa}. FERPA regulates information sharing by requiring that the institution gets explicit affirmative consent to share data with third parties that fall outside listed exceptions. If the documentation a platform provides does not concretely describe how it shares user data with partners or advertisers, it may run afoul of the regulation. 

FERPA only applies to organizations that receive federal funds under certain educational programs. It is mainly enforced when the Department of Education determines that a school or university is in violation. The department then enforces FERPA by withholding federal funds until they come back into compliance.

FERPA specifies what student information can be shared with whom, distinguishing between situations in which consent is required, and those in which it is not. The act permits schools to disclose records to contractors under certain conditions: third-parties must be under the educational institution's direct control, and must be designated as school officials having ``legitimate educational interests.''

FERPA's other notable allowance of data sharing is for directory information---``name, address, telephone number, date and place of birth, honors and awards, and dates of attendance''---which can be shared so long as adequate notice and opportunity to request non-disclosure is provided~\cite{ferpa}. Note that Universities must maintain records of when they share directory information sharing. In all other instances, explicit consent is required, corresponding to a norm that privileges student privacy without informed consent. FERPA reinforces this norm throughout its provisions on disclosure and consent.

While broad in scope, FERPA is limited to a general set of expectations that addresses specific categories of \textit{information types} and \textit{information recipients}. But the rules and formal norms do not translate well to the digital environment, and do not specify transmission principles---when transmitting data is appropriate or who are permissible senders or recipients of data transmissions~\cite{zeide2018learner}. In particular, FERPA does not supply any specific guidance about what educational technology platforms can do with the data they generate and collect about students. However as ``designated school officials'' they may only share that data with other such officials. Other third-party sharing is not permissible, with obligations and limits specified in Department of Education Guidance, originally drafted under the Obama administration and currently applied under the Biden administration~\cite{thirdparty}.

\minihead{State Regulations} State privacy legislation affects platform practices. This includes both general privacy laws (most prominently the California Consumer Privacy Act (CCPA)~\cite{ccpa}) as well as specific laws that regulate student privacy. Specifically, 45 states (which for our purposes includes Washington D.C.) have more than 129 educational privacy laws~\cite{privacycompass,cdtprivacy}, some of which regulate school and student interaction or data collection by digital platforms~\cite{cdtprivacy}.

Many state laws take inspiration from California's Student Online Personal Information Protection Act (SOPIPA)~\cite{fpf} of 2014, which was intended to comprehensively cover K-12 student privacy concerns. In contrast to FERPA, SOPIPA imposed liability on platforms and providers, in addition to schools.

We aggregated 129 state educational privacy laws, as tracked by Student Privacy Compass~\cite{privacycompass} and the Center for Democracy and Technology (CDT)~\cite{cdtprivacy} and coded them to identify and compare information flows and governance patterns, through a combination of manual and hybrid tagging, drawing on established methodologies~\cite{shvartzshnaider2019going,crawfordostrom1995}. We present summary data and publish the corpus alongside this work at \url{https://github.com/edtech-corpus/corpus}.

Almost all states in the corpus had laws that required significant transparency about data sharing practices. 5 states allowed students and families to opt-out of personal information sharing across the board without making a case-by-case determination. 11 states require affirmative consent, opting-in, to share some categories of or all personal information with any recipients outside the school district. 21 state laws included bans on targeted advertising. 6 states were not present in our data set as they had not passed any applicable student privacy laws.


\subsubsection{Analysis}
Regulation at the state level is often more precise than FERPA in addressing specific aspects of digital information flows, limiting platforms as information senders and recipients, as well as articulating clearer transmission principles. Although state level regulations tend to specify more details with respect to permitted information flows in and out of the respective platforms, this layer of governance varies across states and places the burden of compliance on universities rather than the platforms. Moreover, these state laws were primarily designed for the the K-12 context, leaving substantial gaps in the regulatory framework. This mode of privacy regulation imposes more relevant institutions, but is still limited in pertaining to a subset of relevant information subjects and varies significantly from place to place, with schools and universities bearing the burden of compliance, rather than providing a common floor for minimum protection.

A more pervasive issue underlying state or federal laws is that they have limited enforcement mechanisms or penalties for misconduct. For example, under state laws there are no ``private rights of action'', meaning that the laws did not grant students or their guardians the right to sue if a provider violates the law. Primary enforcement is left to state attorneys general, who have limited resources to pursue breaches. Thus, there are few incentives to police compliance with state legal requirements.

As FERPA does not regulate how platforms use data, focusing instead on schools and universities, platforms used in higher education have leeway to use and abuse educational data once it enters their custody.

Universities can fill gaps by introducing their own polices and rules, as well as by extracting binding commitments from commercial partners through contracts. We explore use of these binding commitments in \cref{sec:DPAs}.

\subsection{Privacy Policies}
\label{sec:privacy:policies}

Platforms self-regulate through self-imposed privacy policies, in which they structure and disclose sharing with third parties. Privacy policies may restrict information flows while containing broad language that hedges on specifics. Common sources of flows to third parties include integration between platforms or sharing of data for analytics and marketing.

We manually coded privacy policies for 23 platforms, which corresponded to 18 integrated policies as shown in \cref{tab:privacy_platforms}.

\begin{table}
\small
    \begin{multicols}{2}
    {  \renewcommand{\labelitemi}{\textendash}
    \begin{itemize}
        \item Apple Classroom \\
        Apple Facetime \\
        Apple Schoolwork
        \item BigBlueButton
        \item Blackboard
        \item Blackboard Collaborate
        \item BlueJeans
        \item Canvas
        \item Jitsi
        \item G Suite for Education \\
        Google Classroom
        \item Google Hangouts \\
        Google Meet
        \item GoToMeeting
        \item Microsoft Teams \\
        Microsoft Skype
        \item Microsoft Skype for Business
        \item Panopto
        \item Piazza
        \item Slack
        \item WebEx Meetings
        \item Zoho Meeting
        \item Zoom
    \end{itemize}
    }
    \end{multicols}
    \caption{The 23 platforms whose policies we examined. There were fewer policies than products, as some firms (such as Microsoft) have monolithic policies that apply to groups of products.}
    \label{tab:privacy_platforms}
\end{table}

\begin{table}[t]
\centering
\begin{tabular}{l|c@{}c}
Description                                           & \multicolumn{2}{l}{Frequency} \\ \hline
\textbf{Third Party Sharing}                          &           &                \\
Burden on users to monitor third-parties   & 8 &(44\%)         \\
May share personal data with advertisers     & 8 &(44\%)       \\
Bi-directional sharing                                & 6 &(33\%)        \\
May collect personal data from social media & 7 &(38\%)    \\
\textbf{Location Sharing}                             &           &     \\
Explicitly permit location tracking                   & 10 &(55\%)       \\
May share location data with third-parties            & 4 &(22\%)        \\
Collect location data outside device-provided         & 5 &(27\%)       \\
\end{tabular}
\caption{We identified the privacy practices of 23 platforms from 18 different privacy policies. There are fewer policies than platforms as products owned by a common firm typically shared a policy, indicated by a single bullet spanning multiple platforms.}
\label{tab:priv:policy_findings}
\end{table}

Of the platforms, 13 were mainly available as enterprise products, typically requiring institutional support, and 10 could be adopted at will by individual instructors.

We followed the methodology in~\cite{shvartzshnaider2019going} to annotate statements in each policy based on CI parameters to classify and describe information flows between users, platforms, and third-party entities. For example, in the following quote from Zoho's privacy policy:

\begin{quote}
``We collect information about you only if we need the information for some legitimate purpose.''
\end{quote}

the pronouns ``We'' and ``you'' are  labeled as Receiver (the service provider) and Subject (user) of ``information'', respectively. We label ``for some legitimate purpose'' as transmission principle, i.e., the condition under which the information is being transferred. Note that the statement does not specify the sender of the information. 

The policies we collected serve both as a source of empirical information about patterns in platform practices and a series of case studies that reflect differing governance practices. We summarize our results in \cref{tab:priv:policy_findings}, and present expanded results in \cref{app:sec:privacy-analysis}.

\minihead{Third Party Sharing}
Eight of 18 platform policies explicitly informed users that the burden was on the user to monitor third party firms whose products were integrated with the primary platform. Three platforms specified that agreements with third-party providers provided some privacy protections. The remaining 11 were unclear about third party sharing.

Where a policy applied to EU citizens, the text would typically specify that third parties were also bound to the protections offered by the platform.

While sharing an ID may seem innocuous, student IDs in have long served multiple functions, many of them security sensitive. Blackboard's integration policy permitted Blackboard to share school-provided student IDs with partners.

Eight of 18 platform policies allow the platform to share personal information with advertisers and marketers. Three of 18 policies contained inconclusive language.
Only 2 platforms did not share personal data with third parties for any purpose other than those mandated by law.

Six policies allowed bi-directional sharing. For example, BlueJeans' policy permits collection of user data \textit{from} ``other Service users, third-party service providers...resellers, distributors, your employer, your administrator, publicly available sources, data enrichment vendors, payment and delivery service vendors, advertising networks, analytics providers, and our business partners''---a list that incorporates any conceivable third party.

Seven policies allowed platforms to collect user information from social media, with Zoho going even noting that ``once collected, this information may remain with us even if you delete it from the social media sites.''

Slack's policy, like those of many platforms, places the burden on users to ``check the permissions, privacy settings, and notices for... third-party Services`` whom Slack may receive data from, and to ``contact [Services] for any questions.''

We found significant variation in the level of detail among privacy policies, with only a minority of policies offering detailing specifically when, to whom, and under what conditions information is shared.

\minihead{Location Sharing}
Of the 18 policies we evaluated, 12 policies permitted location tracking, 5 explicitly stated they did not track location, and 1 was unclear.

Of the 12 that collected location data, 4 policies allowed data sharing with third parties. Reasons for sharing and uses permitted varied from the relatively benign (sharing to a mapping company for displaying maps) to the worrisome (sharing for marketing and advertising). Other policies provided broad discretion for uses of anonymized location data. Among the policies with broad language were Google and Apple, whose policy allowed them to share location data with ``partners and licensees to provide and improve location-based products and services.''  

Six of 18 policies mentioned capturing location data using mechanisms other than mobile-device provided location. Notable examples were Slack, which approximated location using information gathered from third parties, and Google, which referenced search data. Four policies did not disclose how they implemented location tracking.

Many policies did not clearly explain why they collected location data, beyond minimal examples under the umbrella category ``improving our services.'' Moreover, the language of the privacy policies could encompass uses that instructors and students might object to---such as Piazza's policy which permits uses ``as required or permitted by law.''

\subsubsection{Analysis}
Our results show that the the governance of platforms and the needs of our stakeholders are not aligned by default. For that reason, considerable negotiation or governance is necessary at the university level to platforms' behavior with our stakeholders' needs.

Our finding that some platforms use a broader range of tracking techniques beyond device-provided location services strips choice away from users. By bypassing device-based restrictions on obtaining location data, platforms subvert users' expectations.

As advertising and marketing third parties are integral parts of the digital economy it is unsurprising that many apps we examined interact with third-parties for marketing or advertising purposes. Policies often failed to enumerate the categories that constituted personal information, leaving platforms with broad discretion to what is appropriate to share.
 
One notable finding was that platform sharing with third parties was in some instances bidirectional---platforms received user data from social networks and other parties, while at the same time transmitting user data to these parties. Platforms may thus be able to build profiles of their users in ways that violate student and institutional expectations, which are derived without this knowledge.

Privacy policies generally reflect defaults applied to individually licensed versions of these tools (which are free or low-cost), reflecting norms in the sense of Lessig's model for governance~\cite{haynes2016,lessig}. Institutions can negotiate provisions when they engage contractually with platforms. But, when individual instructors use these platforms, they do not always realize that free or default licenses do not meet regulatory or normative expectations for privacy protections.


\subsection{Data Protection Addenda}
\label{sec:DPAs}
By leveraging their status as large organizations, universities can negotiate commitments with platforms, called Data Protection Addenda (DPAs), that specify local rules and characterize additional responsibilities and expectations for institutionally supported platforms. 

The DPAs we analyzed reflect three distinct types of contractual relationships: one-to-one, one platform to many universities, and one university to many platforms. Platforms offer template DPAs to make it easier for enterprises, including hospitals and universities, to adopt a given platform. Zoom provides their own templates, emphasizing FERPA and HIPAA obligations, as do Microsoft Teams, Google Hangouts, and Skype for Business. Other types of DPAs include those negotiated between specific universities and specific platforms and those drafted by individual universities and applicable to all vendors. The differences between public universities that negotiate their own agreements and those that use templates are not obviously correlated with factors such as endowment or student body size. 

We coded 50 publicly-available DPAs from a cross section of 41 public universities and 4 private universities. Many universities in our dataset also appear to have other non-public agreements, including with some of the same platforms, as described in the DPAs analyzed and in public FAQs. 



Eleven of the 50 agreements negotiated different sets of rules for educational, organizational, human subject research, and medical uses (relative to university hospital use) within the same document. 10 of these 11 specifically differentiate between educational or enterprise media data and additional protections or scrutiny for university hospital data, such as the documents negotiated between Zoom and the University of Florida, or WebEx and Iowa State University. Another notable modification was Zoom's commitment to allow the University of Illinois to self-host the platform.

4 DPAs for Zoom and 2 DPAs for WebEx were consistent across 6 different universities, including the University of Minnesota and the University of Pittsburgh, implying the platforms' suggested DPAs were employed. In contrast, 18 universities had unique DPAs that correspond with multiple platforms and vendors. 
We also found significant variation in access to data and duration of data retention. For example, under the DPA between Zoom and the University of Virginia, Zoom's obligations \blockquote{survive termination...until all University Data has been returned or Securely Destroyed}. 

The DPAs negotiated by the University of California exhibit similarity as they are designed to comply with the University's Electronic Communications Policy~\cite{ucecp}, showing the impact of local policy. The same constraints can be seen in the DPAs negotiated by Florida State University, which employs information classification guidelines drawn from Florida's public record laws. As a result FSU's agreements include consistent language and requirements for all platforms through which student data is collected, stored, or processed~\cite{FSU}.  This shows how state regulations can shape behavior, even when they don't apply directly.

For the University of Connecticut (UConn), a DPA negotiated by the state with a platform applies not only to the university but to all other public institutions. Besides restraining platforms' practices, rules can impact how universities select features and defaults. In the case of UConn and Zoom, the agreement also places expectations on users (such as obligations to use a passcode and regularly update software), which are enforced by platform settings~\cite{UConn}.

\section{Security Analysis}
\label{sec:security}



\begin{table}[t]
        \centering
        \begin{tabular}{ll}
                \rowcolor{tableheadcolor} Software & Version               \\ \midtopline
                \topmidheader{2}{\centering Windows/macOS}
                Zoom                              & 4.6.10                \\
                Slack                             & 4.5.0 (64bit) / 4.4.2 \\
                BlueJeans                         & 2.19.791 / 2.19.2.128 \\
                Jitsi                             & 2.10.5550             \\
                Cisco WebEx Meetings              & 40.2.16.14            \\
                Cisco WebEx Teams                 & 1.0.0.2               \\
                Microsoft Teams                   & 1.3.0.8663            \\
        \end{tabular}
        \caption{\textbf{Software Evaluated in this study (Desktop Applications)} We evaluate a set of commonly used platforms in remote learning environments. Separate Windows/macOS version numbers are given, where necessary, throughout this work.}
        \label{tab:platforms}
\end{table}

We address the paucity of existing security analyses by performing a deeper analysis of the top video-conferencing and collaboration tools, limiting this portion of our analysis to desktop software (shown in \cref{tab:platforms}). We additionally provide a short analysis of mobile app permissions in \cref{app:security:permissions}.

Our analysis spans four metrics, chosen to maximize ease for an administrator to replicate our procedures.


\begin{table*}[t]
    \begin{tabular}{lccccccc}
        \rowcolor{tableheadcolor}
                       & \textbf{Zoom}   & \textbf{Slack}  & \textbf{BlueJeans} & \textbf{Jitsi}  & \textbf{WebEx (M)} & \textbf{WebEx (T)} & \textbf{MS Teams} \\ \midtopline
        \topmidheader{8}{\centering Windows/MacOS}
        Arch           & i386/AMD64      & AMD64           & AMD64              & AMD64           & i386/AMD64         & AMD64  /    AMD64  & AMD64             \\
        SafeSEH        & \xmark          & N/A             & N/A                & N/A             & \cmark             & N/A                & N/A               \\
        DEP/NX             & \cmark / \cmark & \cmark / \cmark & \cmark / \cmark    & \xmark / \cmark & \cmark / \cmark    & \cmark / \cmark    & \cmark / \cmark   \\
        ASLR           & Low / \cmark     & High / \cmark    & High / \cmark       & \xmark          & Low / \cmark        & High / \cmark       & High / \cmark      \\
        CFI            & \xmark          & \cmark          & \xmark             & \xmark          & \xmark             & \xmark             & \cmark            \\
        Code Signing   & \cmark / \cmark          & \cmark / \cmark         & \cmark  / \cmark           & \cmark / \cmark         & \cmark  / \cmark           & \cmark  / \cmark           & \cmark / \cmark           \\
        Stack Canaries & \cmark / \cmark & \cmark / \cmark & \cmark / \cmark    & \xmark / \cmark & \xmark / \cmark    & \xmark / \cmark    & \xmark / \cmark   \\
    \end{tabular}
    \centering
    \caption{Security features present in end-user binary software on Windows and Mac desktop OSes. Where a feature is available for both Windows and macOS, support in the for the feature is marked for Windows/Mac on the left and right of the slash respectively. Low/High ASLR entropy indicates whether compile time flags necessary for high entropy ASLR were present. Rows with only a single element per entry represent features specific to Windows. As SafeSEH applied only to 32-bit binaries, 64-bit binaries have their respective entries marked N/A. WebEx Teams/Meetings are distinguished by (T) and (M) respectively.}
    \label{fig:exec-security}
\end{table*}

\subsection{Network Traffic Analysis}
\label{sec:network}

We captured network traffic to and from desktop software packages from each platform to determine with whom the application communicated and whether and how it encrypted these flows. We used application-layer traffic analysis in two ways: to search for any qualitative security red-flags (such as obvious failures to encrypt data-in-transit, poor choice of TLS cipher suites), and to look for flows to third parties.

We performed all captures on a clean Windows install using the same software packages profiled in our binary analysis (\cref{sec:binarysec}). We used a monster-in-the-middle attack to interpose between the software packages and the servers they were contacting, allowing us to see unencrypted traffic.

We began each capture, then started and logged in to the software being tested, began a video call with a second client (for all applicable platforms), terminated the call, then terminated the capture. We counted the number of third-party domains to which each connected and looked for domains with no apparent connection to the provision of platforms' services. We also ran the Qualys SSL Labs tests~\cite{qualys-labs} against servers to which client software connected. We excluded all domains that we could identify with the platform operator (e.g.; \url{slack-edge.com}).

We included third-parties that may add platform functionality, such as Gravatar (graphic avatars) or Amazon Chime (video conferencing). Though use of these services may often be justifiable, including them in our results contributes to an understanding of how broadly platforms may share user data.

\subsubsection*{Results}
Only BlueJeans and Slack connected to third party hostnames. BlueJeans connected to New Relic, Microsoft and MixPanel, all for analytics. Slack connected to Gravatar and Amazon Chime, which Slack uses for video calls. The full set of domains are given in \cref{app:security:network}.

Slack, WebEx Teams, and Microsoft Teams all used certificate pinning to verify the identity of the servers they connect to, providing protection against TLS monster-in-the-middle attacks. WebEx Meetings and Zoom presented warnings for untrusted certificates, but allowed users to click through the warning dialogs. If a user clicks through the Zoom warning, Zoom will persistently trust the certificate across executions.

Client software generally requested safe TLS cipher suites, as classified by SSL Labs, but unfortunately all the platforms maintained support for RSA suites, which are known to be weak and vulnerable to attack. Bluejeans and Jitsi supported finite-field Diffie-Hellman cipher suites that researchers similarly warn against~\cite{valenta}. The cipher suites offered by platforms' corresponding servers deviated from from those of their clients, which is intriguing as---had significant thought gone into the choice of suites---the providers would have been able to ensure close matches.

We profiled servers against SSL Labs and found that all platforms' servers received scores of $\mathsf{A}$ or higher for all platforms except for Bluejeans, which received a $\mathsf{B}$ for weak cipher suites. Jitsi requires users to host their own servers and does not provide any, and so was excluded from this analysis.

\subsection{Binary Security}
\label{sec:binarysec}

We evaluate desktop software packages of platforms by building on the \textit{Safety Feature} evaluation criteria of Cyber Independent Testing Labs (Cyber ITL)~\cite{cyberitl}, a nonprofit research organization that attempts to provide consumer friendly security analysis of software and devices. Their approach aims to measure the difficulty ``for an attacker to find a new exploit'' in a given piece of software. 
None of these \textit{features} impose substantial performance penalties and their absence is therefore better explained by ignorance or lack of investment in security.

The full descriptions of the features we analyze are provided in \cref{app:security:binary}.

\subsubsection*{Results} We extracted relevant fields from the first loaded binary image from each software package using tools provided by Microsoft/Apple where available. Where unavailable or when searching for Stack Canaries, we reverse-engineered the binaries by hand. We present our results in \cref{fig:exec-security}.

\minihead{Limitations} Although the results for Jitsi appear below-par compared to the other applications, they are partially an artifact of our methodology. Jitsi is mainly written in Java, except for the main binary which uses native code to initialize the Java Virtual Machine (JVM). Our methodology reveals only the properties of this launcher and not of the underlying JVM present on the users’ system. The Oracle JVM (the predominant instantiation) has well-studied security properties and protections beyond the launcher's. We therefore do not consider Jitsi's results to suggest overall poor security.

\subsection{Known Vulnerabilities and Bug Bounties}
\label{sec:security:cves}
Reports of software failure or flaws in the past predict failure in the future~\cite{gegick2008predictive,bullough2017predicting}. We therefore analyze publicly disclosed platform vulnerabilities. We also collate and discuss the platforms' public \glspl{vdp}, which are an important mechanism to aid firms in detecting and remediating software security flaws~\cite{empiricalbugs}.

\minihead{Vulnerability Disclosure Programs}
Often known as `bug bounty' programs, \glspl{vdp} provide a mechanism for participants to submit flaws to a platform security team, which then (often) fixes the flaw. The programs generally offer rewards for participants---fame, fortune, or both. 

Zoom and Slack both outsource their \glspl{vdp} to HackerOne~\cite{slackvdp,zoomhackerone}, a for-profit operator that has faced criticism for its use of non-disclosure agreements that limit when a reporter may disclose the existence of vulnerabilities \cite{cso_hackerone}. Zoom excluded from its program many types of potential issues including `Attacks requiring MITM' and `Any activity that could lead to the disruption of our service (DoS)'. Following criticism, Zoom hired external consultants to revamp its \gls{vdp}~\cite{lutasecurity}, a process that concluded in July 2020. While Slack has a similar list of exclusions to Zoom, Slack marked them as `\textit{unlikely} to be eligible,' leaving room for discretion.

BlueJeans outsources its \gls{vdp} to Bugcrowd~\cite{bluejeansvdp}. The only significant limitations to its rules-of-engagement are denial of service attacks, and attacks on physical infrastructure or persons. Bluejeans also provides a mechanism for testers to obtain enterprise accounts.

Finally, Cisco and Microsoft retain in-house \glspl{psirt} that handle disclosure. Cisco explicitly includes high-impact vulnerabilities in \emph{third-party} libraries used by their products in their \gls{vdp}~\cite{ciscovdp}.

\minihead{Known Vulnerabilities}
When a vulnerability in software is publicly identified, it is often assigned a number according to the \gls{cve} system. A \gls{cve} ID is stored along with details of the vulnerability in the \gls{nvd}, a separate but related program administered by the National Institute of Standards and Technology. The \gls{nvd} listing includes numerical scores from zero to ten for the impact of the vulnerability when exploited, and for the ease with which the vulnerability can be exploited. These scores as calculated according to the \gls{cvss}.

\begin{figure}[t]
        \input{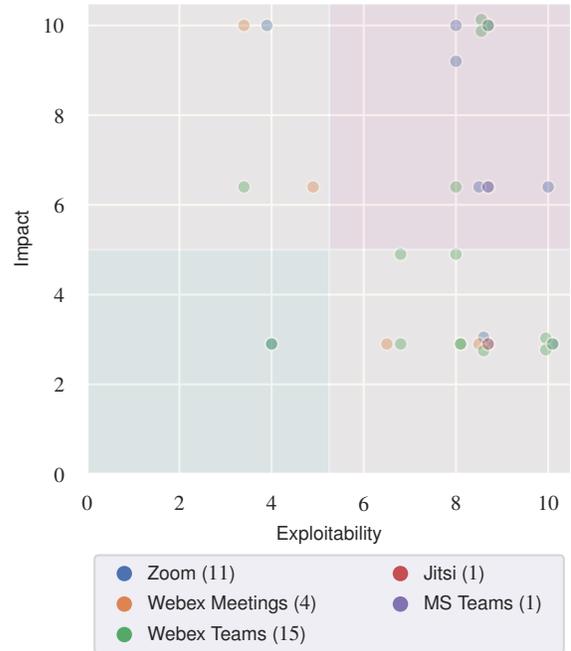}
        \caption{Exploitability and Impact scores for CVEs reported Jan 2010-May 2020. Circular clusters indicate CVEs with the same scores, with markers adjusted for visibility. Regions are shaded to indicate low/high exploitability and impact. Numbers in parentheses indicate the number of CVEs found for a particular software. No CVEs were found for Slack or Bluejeans. {($n = 32$)}}
        \label{fig:cves}
\end{figure}

In \cref{fig:cves} we depict the \gls{cvss}v2 impact and exploitability scores for \glspl{cve} tied to the software we evaluated. While we intended to aggregate all \glspl{cve} from 2010 onward, the earliest \gls{cve} we found for our dataset was issued in 2014---indicative of the relative newness of the platforms evaluated. The vulnerabilities are clustered toward the high-exploitability region of the figure, however this is unsurprising given that the mean impact and exploitability scores across all \glspl{cve} reported since 2010 were 8.04 and 5.04 respectively.

Zoom has many recent CVEs (11). While intense recent attention is no doubt a contributing factor, the substantial number of recent vulnerabilities suggests a systemic component to Zoom's security issues. Further, as our evaluation postdated Zoom's efforts to remediate the aforementioned security issues our results likely understate recent problems with the software. On the other hand, Zoom's rapid improvement in both software and process (that represented a response to disfavorable media coverage) point to a positive trajectory for Zoom.

While Zoom garnered the bulk of negative media attention, in the 2019 alone, five CVEs were issued for WebEx Teams, two of which were high severity each scoring 8.6 on exploitability and 10 on impact. WebEx was issued 15 CVEs total in the reporting period, the most of all software.

\minihead{Limitations}
As many bugs are discovered internally, it is common for vulnerabilities \emph{not} to be assigned a \gls{cve} ID, limiting the extent to which the number of \glspl{cve} for software can be used as a proxy for security. 
For example, during this study Jitsi (which only had one CVE) issued a security bulletin identifying high-severity remote execution bugs in the client software~\cite{jitsibulletin}.
Similarly, in August 2020 HackerOne published a \textit{critical severity} bug in Slack~\cite{hackeroneslack}. 
As of writing, these bugs were not assigned CVEs, despite their significance.

The relative number of CVEs found by researchers may also be more reflective of the scrutiny that has been applied to the platform, more-so than the quality of its software as compared to its peers.

Further, CVEs allocated to software do not reflect security issues with third-party components that developers may include and interface with. A more complete audit of would analyze all included components and vulnerabilities that may be present therein. Exemplifying this concern, in April 2020, Schroder~\cite{schroderzoom} found that Zoom was using outdated libraries that contained known vulnerabilities. Flaws of this type are not reflected in our data.
\section{Related Work}
\label{sec:related}
COVID-19 has led to a flurry of small scale remote learning and working software privacy evaluations in non-academic contexts, with notable efforts from advocacy organizations~\cite{efftools,mozillavideo}. Zoom has been a particular focus of many such evaluations~\cite{citizenlabzoom,schroderzoom,defconzoom,zoomrapid7}. Concurrent to this work, the National Security Agency (NSA) published a guide to and assessment of remote collaboration software for U.S. government employees evaluating many of the same tools~\cite{nsaguide}. Their report relied on product specifications and limited technical observations. Of all products, Zoom has come under particular scrutiny from security researchers who discovered that, contrary to its marketing materials~\cite{zoomwhitepaper}, it used insecure cipher modes~\cite{citizenlabzoom}, did not support end-to-end encryption, and routed users to key-servers in China~\cite{citizenlabzoom, zoomrapid7}. In response to negative media attention Zoom appears to have remediated these issues.

Prior to the COVID-19 pandemic, several academic efforts have examined the security and privacy implications of technologies deployed in education context, as well as the complexity of educational technology procurement~\cite{morrison2014fostering}. In~\cite{weller2018twenty}, Weller provides a historical narrative charting developments in ed-tech, noting that much of the development is due to instructors adopting broader technologies, rather than purpose-built innovation. Balash \textit{et al.} provide the only such evaluation conducted during the pandemic, with a focus on remote proctoring~\cite{balash-proctoring}.

The uptake of ed-tech in schools before the pandemic was on the rise. Gray~\textit{et al.}~\cite{publicschools} performed a comprehensive survey of ed-tech in U.S public schools, finding that as of 2010 between 13\% of teachers regularly used video conferencing technology in the classroom (Tab. 3) and 44\% used software based testing tools (Tab. 6). They also found that 39\% of students accessed a teacher's or course's web-based resource on at least one occasion (Tab. 8). 

A number of prior work evaluated potential privacy harms associated with ed-tech. 

Kelly~\textit{et al.}~\cite{kelly2018state} aggregate and assess 100 privacy policies from purpose-built ed-tech products used in K-12 schools. The authors use this analysis to build a scoring system for `Common Sense', a student privacy advocacy organization. However, they do not evaluate products designed for general use.

Notably, CI guides multiple privacy implications analyses of technologies in the educational ecosystem. Rubel and Jones~\cite{rubel2016student} highlight privacy harms associated with learning analytics in higher education. Zeide and Nissenbaum~\cite{zeide2018learner} show that data-collection driven information handling practices by Massive Open Online Course (MOOC) platforms and Virtual Education providers violate established norms of traditional education settings. 
Jones~{\text et al.}~\cite{jones2020we} reinforce this notion, showing that learning analytics technologies present ``very real challenges to intellectual privacy and contextual integrity'' and that ``colleges and universities need to make concerted efforts to reestablish normative alignment in concert with student expectations.''

Regan and Jesse~\cite{regan2019ethical} explore challenges in applying the oft-vaunted ``Fair Information Practice Principles'' privacy framework~\cite{ware1973records,federal2007fair} to ed-tech. Their concerns center on effects of big-data, autonomy with respect to young people, and surveillance techniques used for purported educational gains. In another work, Regan and Bailey~\cite{regan2019big} find that education focused journals and magazines largely neglect to cover privacy implications of promoted technologies. Both studies show the contextual specificity of educational privacy concerns and the insufficient governance to meet students' expectations in the context of rapid technological change in education.

Peterson~\cite{peterson2016edtech} evaluates federal law's failure to adequately protect student privacy in ed-tech, finding that California's attempts to `band-aid' the gaps highlights the need for overall reform. Given that governance of privacy in the U.S. is highly polycentric, structural governance theory that is compatible with privacy frameworks, such as CI, is useful to understand regulatory and contractual requirements as rules, social expectations as norms, and strategies that bridge gaps and meet local concerns. The institutional grammar, developed by Crawford and Ostrom~\cite{crawfordostrom1995}, and embedded within the governing knowledge commons (GKC) framework~\cite{frischmann2014governing}, is one such approach scholars have employed to study diverse gaps between privacy governance and practice~\cite{sanfilippo2018commons}.

\section{Limitations and Future Work}
\label{sec:future}
Future work can expand the survey to include the expectations of students and other relevant stakeholders with remote learning platforms. Our work is also limited to U.S. universities and regulations. We can learn from the remote learning experiences in other jurisdictions, including the effectiveness of alternative regulatory structures. Indeed, we have seen how regulations in Europe affect U.S. institutions by requiring compliance with the General Data Protection Regulation (GDPR) for students who are accessing virtual classes from a location in the European Union~\cite{fpfgdpr}. Finally, future work can explore the different governance models at universities for procuring and administrating remote learning platforms, and assess how well they address the concerns of stakeholders.
\section{Conclusion and Recommendations}
\label{sec:discussion}

While our results reveal substantial gaps between norms, markets, regulations, and architecture for remote learning platforms, we emphasize that these gaps are not immutable characteristics of the platforms, but rather they reflect issues with default features, settings, and policies to which institutions can negotiate modifications. Our work suggests that DPAs and institutional policies can make platforms modify their default practices that are in tension with institutional values. This approach gives universities the ability to adapt to user expectations---whether for privacy, security, features, usability, or accessibility---and  institutionalize these expectations through the negotiation process.

In other words, universities can use their internal policies to bridge gaps between local needs, community expectations, existing regulation, and practice. Accordingly, we recommend that universities use community privacy norms to set the baseline for privacy strategies and practices. By respecting norms and addressing usability concerns, universities can improve the educational experience and reduce the number of instructors who workaround supported platforms or defaults. 

Crucially, universities do not need to undertake a complex vetting process before licensing software. 
Instead, we recommend IT administrators establish clear principles for how software should respect the norms of the educational context and require developers to offer products that let them customize the software for that setting. Software developers should commit to use that feedback to continually improve the services. We know that significant user issues surface during software use, especially as platforms' functions or uses creep or are employed in new contexts. The key is to build a process to identify concerns or needs and rapidly fix problems, especially privacy harms, including thwarted expectations, control, and informed choice~\cite{citron2021privacy}. 

\smallskip
We offer the following specific recommendations:

\begin{itemize}
    \item \textbf{Identify User Expectations:}
Administrators should periodically solicit concerns and expectations from instructors and students about the major platforms the universities have, or intend to, license. Our survey revealed that instructors were more likely to respond to questions based on specific cases. At the same time, administrators should be sensitive to how certain design choices (e.g., video recordings) may disproportionately impact vulnerable groups who are often targets of online abuse~\cite{powell-harassment, gardiner-harassment} and should design surveys to surface such concerns.

\item \textbf{Negotiate Specific Practices:}
While platforms may offer education-specific terms in their contracts, universities should negotiate terms based on an individualized needs assessment. Among the changes universities may want to request are options for local hosting, third-party sharing, limiting how platforms use data, and separating the institutions' data from that of other platform users.

\item \textbf{Penalize Noncompliance:}
Regulators and universities should work together to identify instances of noncompliance and create incentives for the platforms to take prompt action to remediate harms, in the inclusive legal sense~\cite{citron2021privacy}. In particular, we recommend strengthening regulations to ensure that software used in educational institutions comply with state and federal laws and that there are mandatory baseline security practices for educational technology that parallel those financial institutions are required to adopt to protect consumer information under the Federal Trade Commission's Safeguards Rule~\cite{safeguardsrule}.

\item \textbf{Popularity Does Not Guarantee Security:}
If institutions fail to adequately prioritize security, platforms will continue to prioritize growth over product improvements. Our evaluation reflects these misaligned incentives, showing little correlation between product security and popularity.

\end{itemize}
The shift to virtual learning requires many sacrifices from instructors and students already---we should mitigate their real harms, not further sacrifice usability, security, and privacy.





\section*{Acknowledgements}
We thank our reviewers for their helpful advice on improving the paper. We are also grateful to members of the NITRD and PLSC for their feedback on drafts of this work.

\bibliographystyle{plain}
\bibliography{paper}

\begin{thebibliography}{10}

\bibitem{bluejeansvdp}
{Bugcrowd - BlueJeans}.
\newblock \url{https://bugcrowd.com/bluejeans}.

\bibitem{slackvdp}
{HackerOne - Slack}.
\newblock \url{https://hackerone.com/slack}.

\bibitem{zoomhackerone}
{HackerOne - Zoom}.
\newblock \url{https://hackerone.com/zoom}.

\bibitem{zoomwhitepaper}
{Security Guide: Zoom Video Communications, Inc}.
\newblock Technical report, {Zoom Inc.}
\newblock
  \url{https://web.archive.org/web/20200403154149/https://zoom.us/docs/doc/Zoom-Security-White-Paper.pdf}.

\bibitem{fpf}
Fpf guide to protecting student data under sopipa.
\newblock Technical report, Future of Privacy Forum, 2016.

\bibitem{FSU}
Information security and privacy standard terms and conditions.
\newblock Technical report, Florida State University, 2018.

\bibitem{jitsibulletin}
{Multiple Remote Code Execution issues}.
\newblock Technical report, Jitsi, 2020.
\newblock
  \url{https://github.com/jitsi/security-advisories/blob/master/advisories/JSA-2020-0001.md}.

\bibitem{nsaguide}
{Selecting and Safely Using Collaboration Services for Telework}.
\newblock Technical report, National Security Agency, 2020.

\bibitem{ferpa}
34~CFR Part~99 20~U.S. Code~§1232g.
\newblock {Family Educational Rights and Privacy Act (FERPA) }.

\bibitem{ccpa}
AB-375.
\newblock {California Consumer Privacy Act of 2018}.

\bibitem{abadi2009control}
Mart{\'\i}n Abadi, Mihai Budiu, {\'U}lfar Erlingsson, and Jay Ligatti.
\newblock Control-flow integrity principles, implementations, and applications.
\newblock {\em ACM Transactions on Information and System Security (TISSEC)},
  13(1):1--40, 2009.

\bibitem{defconzoom}
Mazin Ahmed.
\newblock Hacking zoom: Uncovering tales of security vulnerabilities in zoom,
  2020.

\bibitem{cdtprivacy}
BakerHostetler.
\newblock {State Student Privacy Law Compendium}, 2019.
\newblock
  \url{https://cdt.org/wp-content/uploads/2016/10/CDT-Stu-Priv-Compendium-FNL.pdf}.

\bibitem{balash-proctoring}
David~G. Balash, Dongkun Kim, Darika Shaibekova, Rahel~A. Fainchtein, Micah
  Sherr, and Adam~J. Aviv.
\newblock {Examining the Examiners: Students’ Privacy and Security
  Perceptions of Online Proctoring Services}.
\newblock {\em USENIX Symposium on Usable Privacy and Security}, 2021.

\bibitem{zoomrapid7}
Tod Beardsley.
\newblock {Dispelling Zoom Bugbears: What You Need to Know About the Latest
  Zoom Vulnerabilities}.
\newblock Technical report, Rapid7, 2020.
\newblock
  \url{https://blog.rapid7.com/2020/04/02/dispelling-zoom-bugbears-what-you-need-to-know-about-the-latest-zoom-vulnerabilities/}.

\bibitem{mozillavideo}
Ashley Boyd.
\newblock {Which Video Call Apps Can You Trust?}, 2020.
\newblock
  \url{https://blog.mozilla.org/blog/2020/04/28/which-video-call-apps-can-you-trust/}.

\bibitem{bullough2017predicting}
Benjamin~L Bullough, Anna~K Yanchenko, Christopher~L Smith, and Joseph~R
  Zipkin.
\newblock Predicting exploitation of disclosed software vulnerabilities using
  open-source data.
\newblock In {\em Proceedings of the 3rd ACM on International Workshop on
  Security And Privacy Analytics}, pages 45--53, 2017.

\bibitem{canvas-conference}
{Canvas}.
\newblock {What are Conferences?}, 2021.
\newblock
  \url{https://community.canvaslms.com/t5/Canvas-Basics-Guide/What-are-Conferences/ta-p/53}.

\bibitem{lutasecurity}
Catalin Cimpanu.
\newblock Zoom to revamp bug bounty program, bring in more security experts.
\newblock {\em ZDNet}.
\newblock
  \url{https://www.zdnet.com/article/zoom-to-revamp-bug-bounty-program-bring-in-more-security-experts/}.

\bibitem{ciscovdp}
Cisco.
\newblock {Security Vulnerability Policy}.
\newblock
  \url{https://tools.cisco.com/security/center/resources/security_vulnerability_policy.html}.

\bibitem{citron2021privacy}
Danielle~Keats Citron and Daniel~J Solove.
\newblock Privacy harms.
\newblock {\em Available at SSRN}, 2021.

\bibitem{federal2007fair}
Federal~Trade Commission, Federal~Trade Commission, et~al.
\newblock Fair information practice principles.
\newblock 25, 2007.

\bibitem{privacycompass}
Student~Privacy Compass.
\newblock {STATE STUDENT PRIVACY LAWS}, 2019.
\newblock \url{http://studentprivacycompass.org/state-laws/}.

\bibitem{UConn}
Connecticut.
\newblock Zoom guideance, 2020.
\newblock
  \url{https://portal.ct.gov/Government/Work-from-Home-Technology-Resources/Zoom-Guidance}.

\bibitem{cowan1998stackguard}
Crispan Cowan, Calton Pu, Dave Maier, Jonathan Walpole, Peat Bakke, Steve
  Beattie, Aaron Grier, Perry Wagle, Qian Zhang, and Heather Hinton.
\newblock Stackguard: Automatic adaptive detection and prevention of
  buffer-overflow attacks.
\newblock In {\em USENIX Security Symposium}, volume~98, pages 63--78. San
  Antonio, TX, 1998.

\bibitem{crawfordostrom1995}
Sue~ES Crawford and Elinor Ostrom.
\newblock A grammar of institutions.
\newblock {\em American political science review}, pages 582--600, 1995.

\bibitem{cyberitl}
{Cyber Independent Testing Lab}.
\newblock {Methodology: ``How difficult is it for an attacker to find a new
  exploit for this software?''}.
\newblock Available at: \url{https://cyber-itl.org/about/methodology/}.

\bibitem{privacy-badger}
{Electronic Frontier Foundation}.
\newblock {Privacy Badger}, 2021.
\newblock \url{https://privacybadger.org}.

\bibitem{safeguardsrule}
{Federal Trade Commission}.
\newblock Public workshop examining information security for financial
  institutions and information related to changes to the safeguards rule.
\newblock {\em Federal Register}, 85(45):13082--1308, mar 2020.

\bibitem{freiwald}
Susan Freiwald.
\newblock Comparative institutional analysis in cyberspace: the case of
  intermediary liability for defamation.
\newblock {\em Harv. JL \& Tech.}, 14:569, 2000.

\bibitem{frischmann2014governing}
Brett~M Frischmann, Michael~J Madison, and Katherine~Jo Strandburg.
\newblock {\em Governing knowledge commons}.
\newblock Oxford University Press, 2014.

\bibitem{gardiner-harassment}
Becky Gardiner.
\newblock {''It’s a terrible way to go to work:'' what 70 million readers’
  comments on the Guardian revealed about hostility to women and minorities
  online}.
\newblock {\em Feminist Media Studies}, 18(4):592--608, 2018.

\bibitem{gegick2008predictive}
Michael~C Gegick, Laurie~Ann Williams, and Mladen~A Vouk.
\newblock Predictive models for identifying software components prone to
  failure during security attacks.
\newblock Technical report, North Carolina State University. Dept. of Computer
  Science, 2008.

\bibitem{publicschools}
Lucinda Gray, Nina Thomas, and Laurie Lewis.
\newblock Teachers' use of educational technology in us public schools: 2009.
  first look. nces 2010-040.
\newblock {\em National Center for Education Statistics}, 2010.

\bibitem{haynes2016}
David Haynes, David Bawden, and Lyn Robinson.
\newblock A regulatory model for personal data on social networking services in
  the uk.
\newblock {\em International Journal of Information Management},
  36(6):872--882, 2016.

\bibitem{jones2020we}
Kyle~ML Jones, Andrew Asher, Abigail Goben, Michael~R Perry, Dorothea Salo,
  Kristin~A Briney, and M~Brooke Robertshaw.
\newblock {``We're being tracked at all times'': Student perspectives of their
  privacy in relation to learning analytics in higher education}.
\newblock {\em Journal of the Association for Information Science and
  Technology}, 2020.

\bibitem{kelly2018state}
G~Kelly, J~Graham, and B~Fitzgerald.
\newblock State of edtech privacy report.
\newblock {\em Common Sense Privacy Evaluation Initiative}, 2018.

\bibitem{STRIDE}
Loren Kohnfelder and Praerit Garg.
\newblock The threats to our products.
\newblock Technical report, Microsoft, 1999.
\newblock Internal Microsoft Magazine. Available at
  \url{https://adam.shostack.org/microsoft/The-Threats-To-Our-Products.docx}.

\bibitem{lessig}
Lawrence Lessig.
\newblock {\em Code: And other laws of cyberspace, version 2.0}.
\newblock Basic Books, 2006.

\bibitem{ling2020first}
Chen Ling, Utkucan Balc{\i}, Jeremy Blackburn, and Gianluca Stringhini.
\newblock A first look at zoombombing.
\newblock {\em arXiv preprint arXiv:2009.03822}, 2020.

\bibitem{lock2010}
R.~{Lock} and I.~{Sommerville}.
\newblock Modelling and analysis of socio-technical system of systems.
\newblock In {\em 2010 15th IEEE International Conference on Engineering of
  Complex Computer Systems}, pages 224--232, 2010.

\bibitem{aslrng}
Hector Marco-Gisbert and Ismael Ripoli~Ripoli.
\newblock {Address Space Layout Randomization Next Generation}.
\newblock {\em Applied Sciences}, 2019.

\bibitem{citizenlabzoom}
Bill Marczak and John Scott-Railton.
\newblock {Move Fast and Roll Your Own Crypto: A Quick Look at the
  Confidentiality of Zoom Meetings}.
\newblock {\em CitizenLab}.
\newblock
  \url{https://citizenlab.ca/2020/04/move-fast-roll-your-own-crypto-a-quick-look-at-the-confidentiality-of-zoom-meetings/}.

\bibitem{morrison2014fostering}
Jennifer Morrison, Steven Ross, Roisin Corcoran, and AJ~Reid.
\newblock Fostering market effciency in k--tech procurement.
\newblock Technical report, Johns Hopkins University, 2014.

\bibitem{nissenbaum2009privacy}
Helen Nissenbaum.
\newblock {\em Privacy in context: Technology, policy, and the integrity of
  social life}.
\newblock Stanford University Press, 2009.

\bibitem{efftools}
Lindsay Oliver.
\newblock {What You Should Know About Online Tools During the COVID-19 Crisis},
  2020.
\newblock
  \url{https://www.eff.org/deeplinks/2020/03/what-you-should-know-about-online-tools-during-covid-19-crisis}.

\bibitem{hackeroneslack}
oskarsv.
\newblock {Remote Code Execution in Slack desktop apps + bonus}, 2020.
\newblock \url{https://hackerone.com/reports/783877}.

\bibitem{team2003pax}
{PaX Team}.
\newblock Pax address space layout randomization (aslr).
\newblock 2003.

\bibitem{peterson2016edtech}
Dylan Peterson.
\newblock Edtech and student privacy: California law as a model.
\newblock {\em Berkeley Technology Law Journal}, 31(2):961--996, 2016.

\bibitem{cso_hackerone}
J.M. Porup.
\newblock Bug bounty platforms buy researcher silence, violate labor laws,
  critics say.
\newblock {\em CSO Online}.
\newblock
  \url{https://www.csoonline.com/article/3535888/bug-bounty-platforms-buy-researcher-silence-violate-labor-laws-critics-say.html}.

\bibitem{powell-harassment}
Anastasia Powell, Adrian~J Scott, and Nicola Henry.
\newblock Digital harassment and abuse: Experiences of sexuality and gender
  minority adults.
\newblock {\em European Journal of Criminology}, 17(2):199--223, 2020.

\bibitem{ferpaoverview}
{Privacy Technical Assistance Center}.
\newblock {Protecting Student Privacy While Using Online Educational Services:
  Requirements and Best Practices}, 2014.

\bibitem{thirdparty}
{Privacy Technical Assistance Center}.
\newblock Responsibilities of third-party service providers under ferpa, 2015.

\bibitem{qualys-labs}
{Qualys SSL Labs}.
\newblock {SSL Server Test}, 2021.
\newblock \url{https://www.ssllabs.com/ssltest/}.

\bibitem{regan2019big}
Priscilla~M Regan and Jane Bailey.
\newblock Big data, privacy and education applications.
\newblock {\em Ottawa Faculty of Law Working Paper}, (2019-44), 2019.

\bibitem{regan2019ethical}
Priscilla~M Regan and Jolene Jesse.
\newblock Ethical challenges of edtech, big data and personalized learning:
  twenty-first century student sorting and tracking.
\newblock {\em Ethics and Information Technology}, 21(3):167--179, 2019.

\bibitem{rubel2016student}
Alan Rubel and Kyle~ML Jones.
\newblock Student privacy in learning analytics: An information ethics
  perspective.
\newblock {\em The information society}, 32(2):143--159, 2016.

\bibitem{sanfilippo2018commons}
Madelyn Sanfilippo, Brett Frischman, and Katherine Strandburg.
\newblock Privacy as commons: Case evaluation through the governing knowledge
  commons framework.
\newblock {\em Journal of Information Policy}, 8:116--166, 2018.

\bibitem{sanfilippo2020disaster}
Madelyn~R Sanfilippo, Yan Shvartzshnaider, Irwin Reyes, Helen Nissenbaum, and
  Serge Egelman.
\newblock Disaster privacy/privacy disaster.
\newblock {\em Journal of the Association for Information Science and
  Technology}, 71(9):1002--1014, 2020.

\bibitem{schroderzoom}
Thorsten Schr\"oder.
\newblock {Zoom Endpoint-Security Considerations}.
\newblock Technical report, 2020.
\newblock \url{https://dev.io/posts/zoomzoo/}.

\bibitem{defeatingaslr}
Martin Shudrak.
\newblock {Defeating Windows ASLR via low-entropy shared libraries in 2 hours}.
\newblock Technical report, 2020.
\newblock
  \url{https://medium.com/@mxmssh/defeating-windows-aslr-via-32-bit-shared-libraries-in-2-hours-1e225e182155}.

\bibitem{shvartzshnaider2019going}
Yan Shvartzshnaider, Noah Apthorpe, Nick Feamster, and Helen Nissenbaum.
\newblock Going against the (appropriate) flow: a contextual integrity approach
  to privacy policy analysis.
\newblock In {\em Proceedings of the AAAI Conference on Human Computation and
  Crowdsourcing}, volume~7, pages 162--170, 2019.

\bibitem{shvartzshnaider2016learning}
Yan Shvartzshnaider, Schrasing Tong, Thomas Wies, Paula Kift, Helen Nissenbaum,
  Lakshminarayanan Subramanian, and Prateek Mittal.
\newblock Learning privacy expectations by crowdsourcing contextual
  informational norms.
\newblock In {\em HCOMP}, pages 209--218, 2016.

\bibitem{nytimes-dartmouth}
{Singer, Natasha and Krolik, Aaron}.
\newblock Online cheating charges upend dartmouth medical school, 2021.

\bibitem{sotirov2009bypassing}
Alexander Sotirov.
\newblock Bypassing memory protections: The future of exploitation.
\newblock In {\em USENIX Security}, 2009.

\bibitem{exploringcfg}
Jack Tang.
\newblock Exploring control flow guard in windows 10.
\newblock Technical report, Trend Micro Threat Solution Teahm, 2015.

\bibitem{aslrwindows}
Jacob Thompson.
\newblock {Six Facts about Address Space Layout Randomization on Windows}.
\newblock Technical report, FireEye, 2020.
\newblock
  \url{https://www.fireeye.com/blog/threat-research/2020/03/six-facts-about-address-space-layout-randomization-on-windows.html}.

\bibitem{ucecp}
{University of California Office of the President}.
\newblock {Electronic Communications Policy}, 2005.
\newblock \url{https://policy.ucop.edu/doc/7000470/ElectronicCommunications}.

\bibitem{valenta}
Luke~Taylor Valenta.
\newblock Measuring and securing cryptographic deployments.
\newblock 2019.

\bibitem{empiricalbugs}
T.~{Walshe} and A.~{Simpson}.
\newblock An empirical study of bug bounty programs.
\newblock In {\em 2020 IEEE 2nd International Workshop on Intelligent Bug
  Fixing (IBF)}, pages 35--44, 2020.

\bibitem{ware1973records}
Willis~H Ware.
\newblock Records, computers and the rights of citizens.
\newblock 1973.

\bibitem{weller2018twenty}
Martin Weller.
\newblock Twenty years of edtech.
\newblock {\em Educause Review Online}, 53(4):34--48, 2018.

\bibitem{s4issues}
Christian Wressnegger, Fabian Yamaguchi, Alwin Maier, and Konrad Rieck.
\newblock Twice the bits, twice the trouble: Vulnerabilities induced by
  migrating to 64-bit platforms.
\newblock In {\em Proceedings of the 2016 ACM SIGSAC Conference on Computer and
  Communications Security}, CCS '16, page 541–552, New York, NY, USA, 2016.
  Association for Computing Machinery.

\bibitem{fpfgdpr}
Gabriela Zanfir-Fortuna.
\newblock {The General Data Protection Regulation: Analysis and Guidance for US
  Higher Education Institutions}.
\newblock Technical report, Future of Privacy Forum, 2021.

\bibitem{zeide2018learner}
Elana Zeide and Helen Nissenbaum.
\newblock Learner privacy in moocs and virtual education.
\newblock {\em Theory and Research in Education}, 16(3):280--307, 2018.

\end{thebibliography}

\newpage
\appendix
\section{Additional Survey Details and Results}
\crefalias{section}{appendix}
\label{app:survey:breakdown}

Here, we describe our instructor and administrator surveys in detail.

\minihead{Survey details} The instructor and administrator surveys were hosted on Qualtrics. The surveys were conducted from July 2020 to January 2021. No compensation was given to participants.

The instructor survey begins by asking participants general information about their teaching: their institution, grade level, field, and sizes of classrooms they teach. Next, we ask what video conferencing platforms instructors use (from a multiple choice list), and whether they use a personal or institutional provided version. If instructors use an institution and personal version, they are instructed to mark `institutional'. We then ask instructors the reasons that instructors choose to use each platform. We then ask instructors the same questions for remote learning platforms other than those used for remote conferencing. For all these questions, instructors have the option to include additional platforms not listed in our survey using an `Other' option.

Finally, we ask instructors five freeform questions involving their concerns and expectations with remote learning platforms. First, we ask about their own concerns, followed by concerns they have heard from their students. Next, we ask what features instructors consider essential for a remote learning platform to have. Finally, we ask instructors to discuss privacy features and security features that they currently use or desire in a platform.

In the administrator survey, we first ask administrators what remote learning platforms their institution contracted with or supported before COVID-19 in a multiple-choice list. We then ask questions about the administrators' decision making processes for procuring new platforms at their university. We ask whether their universities have a documented process for selecting teaching tools, and whether they would share this document with us. We also ask whether they follow this process strictly, and to explain why if not.

Next, we ask a series of freeform questions regarding platforms at administrators' universities. We ask what remote teaching platforms administrators considered adding since COVID-19 began; what platforms were added since COVID-19; whether admins are trying to request new features or cancel any platform licenses after COVID-19; and what platforms were rejected after review. Administrators are also asked to explain each answer.

Then, we ask administrators to rate the influence of different first parties and third parties on the platform procurement process from `A great deal' to `None at all'. We ask how administrators consult these parties (survey, interviews, etc.) We also ask admins to rate the importance of different features of remote learning platforms, from `Extremely important' to `Not at all important'.

We ask admins to describe any additional features/protections they negotiated with platforms. We ask whether admins' universities have processes for collecting feedback from instructors/students on platforms currently in use. We also ask which platforms gained users since COVID-19.

We ask admins to rate sources of information by the likelihood of considering them when addressing future platform or policy adoption, from 'Extremely likely' to 'extremely unlikely', and what information would be most helpful for admins to learn. Finally, we allow admins to leave additional comments about remote learning.


\begin{figure}[ht]
        \centering
        \includegraphics[width=\linewidth]{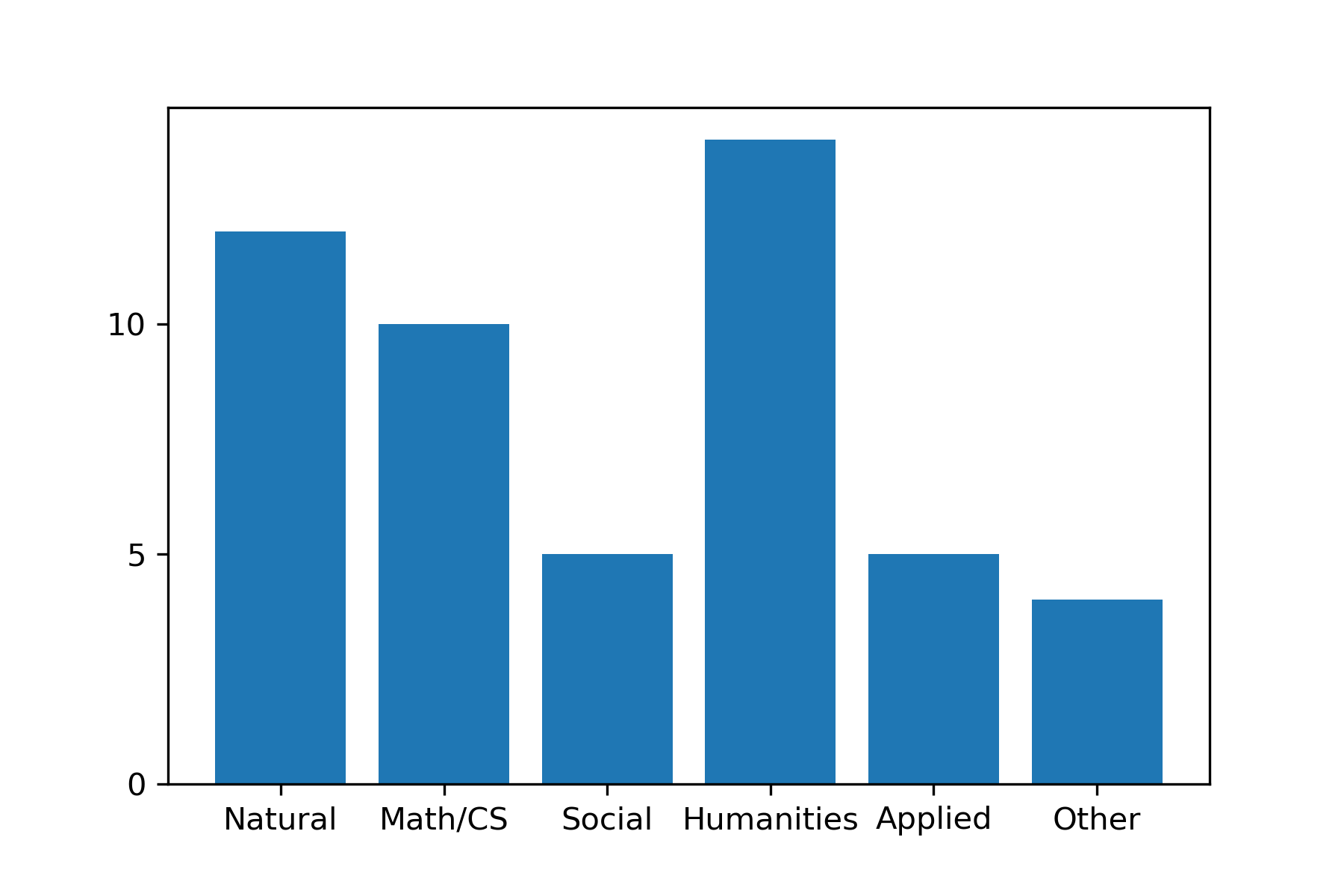}
        \caption{Subjects taught per instructor.}
        \label{fig:subject-taught}
\end{figure}

\begin{figure}[ht]
        \centering
        \includegraphics[width=\linewidth]{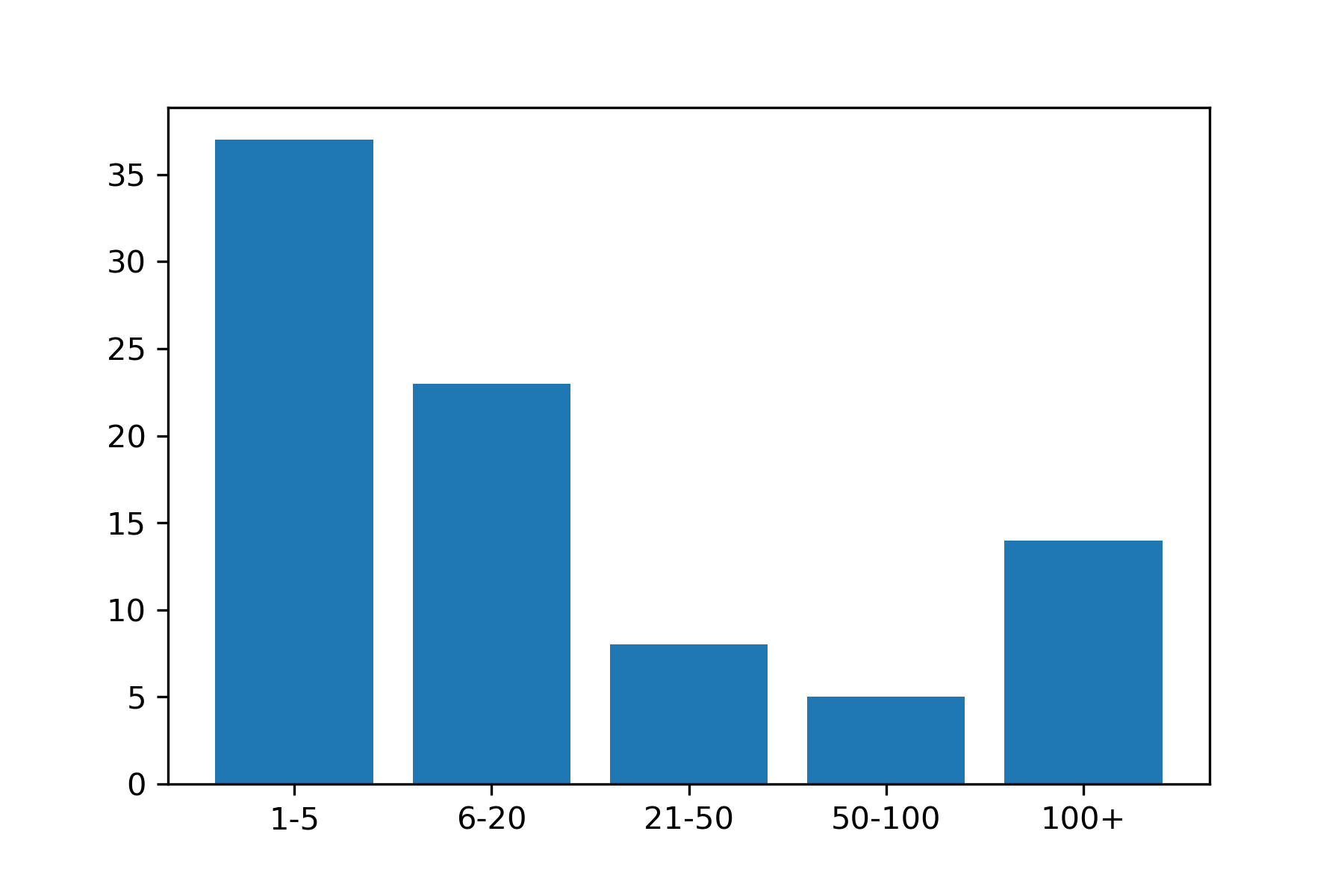}
        \caption{Class video conferencing sizes.}
        \label{fig:class-size}
\end{figure}

\begin{figure*}[ht]
        \centering
        \includegraphics[width=\linewidth]{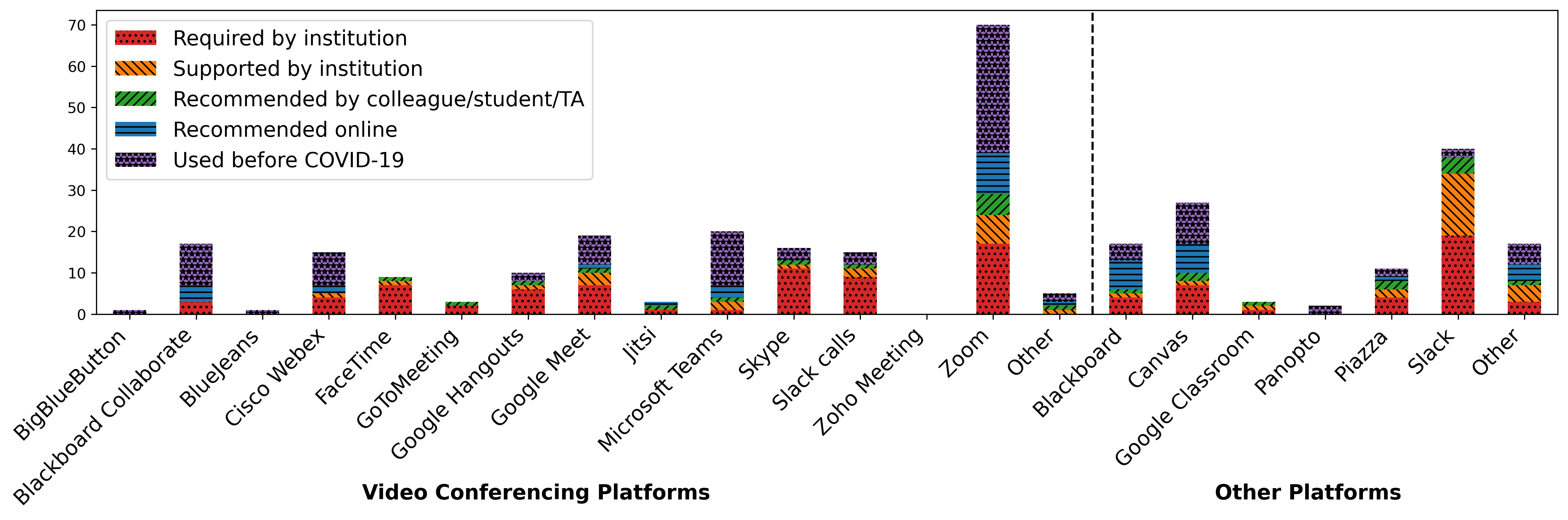}
        \caption{Motivations for instructors using platforms.}
        \label{fig:motivations}
\end{figure*}

\minihead{Results}
We received 128 total responses to our instructor survey. Of our respondents 109 (85.2\%) taught at an undergraduate/graduate level, while the remaining 19 (14.8\%) taught K-12 or at professional schools (dentistry, tech literacy, etc.) Respondents came from a diverse set of disciplines and locations, and taught classes from small discussion groups to 100+ lectures.

We asked instructors about their teaching background: the institution they teach at, their class level and sizes, and discipline area. We then had instructors explain the video conferencing and other remote learning platforms they use, including whether they use a personal or institutional version, as well as their motivations for using each platform. Next, we asked instructors to describe their complaints regarding platforms, and complaints they have received from students and others involved in the course. Finally, we asked instructors what features they consider essential for a remote teaching platform to have---separately for general features, privacy features, and security features.

At least \ninstructors (38\%) of instructors were located in the United States and at least 8 (6.2\%) in Europe, with others in the Middle East, India, and Australia. The others respondents left this field blank, or gave an abbreviation that matched universities in multiple countries.

Note that we limit all results in the main text and below to undergraduate/graduate instructors from U.S. institutions.

\minihead{Class subjects and sizes} \cref{fig:subject-taught} shows the subjects taught by each instructor. \cref{fig:class-size} shows the sizes of the groups that each instructor reported regularly videoconferencing with. Note that we allowed instructors to report multiple sizes of groups, so \cref{fig:class-size} represents the number of conferencing \textit{settings} rather than the number of instructors.

\minihead{Motivations for platforms}
In \cref{fig:motivations}, we present the motivations that instructors reported for using each video conferencing and remote learning platforms. Zoom and Teams were frequently used by instructors prior to COVID-19, while Zoom, Slack and Skype were required by many institutions. These results may highlight the use of tools for communication among other instructors and administrators currently and in the past, in addition to educating students.

\minihead{Complaints from instructors} Next, we asked instructors to describe their frustrations with the platforms they currently use; 28 (57\%) of instructors provided complaints on a wide variety of issues.

Two instructors reported frustration with the fact that no single platform can handle all of their needs, forcing them to ``cobble together'' multiple platforms to run a course, each with its own learning curve for instructors and for students. instructors cited steep learning curves and/or a lack of documentation for understanding how to efficiently configure and use platforms such as Zoom, Blackboard, and Kaltura.

Other instructors reported frustrations with the fact that they were required to use tools they did not like, with one instructor describing being ``forced against \textins{their} will'' to use Zoom. In fact, many complaints were directed towards Zoom. Several instructors wished that Zoom would allow students to move between breakout rooms or communicate across breakout rooms to facilitate better class discussions. One instructor disliked that the ``Zoom chat history is not available for students who join late,'' forcing the instructor to repeatedly paste links for tardy students.

Instructors also found it difficult to engage with students, and one wished they could ``force students to keep their cameras on to monitor their engagement.''

\minihead{Complaints from students} 24 (48.9\%) of instructors also reported complaints from their students about the remote learning platforms they use.

Many students experienced issues with low bandwidth or other glitches that hamper their ability to participate in class. Students felt generally fatigued from conducting class entirely through video conferencing. Students also found it difficult to communicate with other students in the class, and only very small breakout rooms allowed for meaningful engagement.

\section{Privacy Policy Analysis}
\label{app:sec:privacy-analysis}
In \cref{app:tab:third-party-privacy}, we present the breakdown for third-party tracking in platform privacy policies. In \cref{app:tab:location-privacy}, we present the breakdown for location tracking. A \cmark means the policy explicitly permits the activity; an \xmark means the policy explicitly forbids the activity; others do not specify.

Blackboard Collaborate supercedes Blackboard's privacy policy in one case, where Blackboard Collaborate does not share data with third-party advertisers (while Blackboard may).

Google's education policy supercedes Google's main primary policy in one case: Google does not share G Suite for Education data with third-party advertisers except for "non-core" products such as Maps and YouTube.

\begin{table*}[ht]
\centering
\begin{tabular}{lcccc}
\rowcolor{tableheadcolor}\textbf{Third-Party Tracking} &
  Burden on Users to Monitor &
  Shared With Advertisers &
  Bi-directional Sharing &
  Social Media Data \\ \hline
\multicolumn{1}{l|}{Apple} &
   &
  \xmark &
   &
   \\ \hline
\multicolumn{1}{l|}{BigBlueButton} &
  \cmark &
   &
   & \cmark
   \\ \hline
\multicolumn{1}{l|}{Blackboard} &
   &
  \cmark &
   &
  \xmark \\ \hline
\multicolumn{1}{l|}{Blackboard Collaborate} &
   &
  \xmark &
   &
  \xmark \\ \hline
\multicolumn{1}{l|}{BlueJeans} &
  \cmark &
  \cmark &
  \cmark &
  \cmark \\ \hline
\multicolumn{1}{l|}{Canvas} &
   &
  \cmark &
   &
  \cmark \\ \hline
\multicolumn{1}{l|}{Cisco} &
   &
   &
   &
  \cmark \\ \hline
 \multicolumn{1}{l|}{Google} &
   &
 \cmark  &
   &
  \cmark \\ \hline
\multicolumn{1}{l|}{Google Education} &
   &
 \xmark \text{*}  &
   &
  \cmark \\ \hline
\multicolumn{1}{l|}{GoToMeeting/LogMeIn} &
  \cmark &
  \cmark &
  \cmark &
  \xmark \\ \hline
\multicolumn{1}{l|}{Jitsi} &
   &
  \xmark &
   &
   \\ \hline
\multicolumn{1}{l|}{MS Teams/Skype} &
  \cmark &
   &
  \cmark &
   \\ \hline
\multicolumn{1}{l|}{Panopto} &
  \cmark &
   &
  \cmark &
   \\ \hline
\multicolumn{1}{l|}{Piazza} &
  \cmark &
  \cmark &
   &
   \\ \hline
\multicolumn{1}{l|}{Skype for Business} &
   &
   &
   &
   \\ \hline
\multicolumn{1}{l|}{Slack} &
  \cmark &
  \cmark &
  \cmark &
   \\ \hline
\multicolumn{1}{l|}{Zoom} &
   &
  \xmark &
   &
   \\ \hline
\multicolumn{1}{l|}{Zoho} &
  \cmark &
  \cmark &
  \cmark &
  \cmark
\end{tabular}
\caption{The breakdown of third-party data sharing in privacy policies. \text{*} = Google's G Suite Education policy specifies that data is only shared with third parties for non-core services, such as Maps and YouTube.}
\label{app:tab:third-party-privacy}
\end{table*}

\begin{table*}[h!]
\centering
\begin{tabular}{lccccc}
\rowcolor{tableheadcolor} \textbf{Location Tracking} & Permitted &
  Active Tracking &
  Shared with Third Parties &
   Inferred Location &
  Exact Location \\ \hline
\multicolumn{1}{l|}{Apple}   & \cmark               &                          &                           & \cmark &                           \\ \hline
\multicolumn{1}{l|}{BigBlueButton}    &  \xmark    & \xmark                     & \xmark                     & \xmark                     & \xmark                     \\ \hline
\multicolumn{1}{l|}{Blackboard}    &    \xmark     & \xmark                     & \xmark                     & \xmark                     & \xmark                     \\ \hline
\multicolumn{1}{l|}{Blackboard Collaborate}   &  \xmark        & \xmark                     & \xmark                     & \xmark                     & \xmark                     \\ \hline
\multicolumn{1}{l|}{BlueJeans}    & \cmark         &                         & \cmark &                         & \cmark \\ \hline
\multicolumn{1}{l|}{Canvas}         &        & \xmark                         & \xmark                         & \xmark                         &                           \\ \hline
\multicolumn{1}{l|}{Cisco}       & \cmark           & \cmark & \xmark                         & \cmark & \xmark                     \\ \hline
\multicolumn{1}{l|}{Google}       & \cmark          &                         &            \xmark             & \cmark & \cmark \\ \hline
\multicolumn{1}{l|}{Google Education}     & \cmark            &                         &            \xmark             & \cmark & \cmark \\ \hline
\multicolumn{1}{l|}{GoToMeeting/LogMeIn} & \cmark &
   &
  \cmark &
   &
  \cmark \\ \hline
\multicolumn{1}{l|}{Jitsi}                  & \xmark                         & \xmark                         & \xmark                         & \xmark                     \\ \hline
\multicolumn{1}{l|}{MS Teams/Skype} & \cmark &
   &
   &
  \cmark &
  \cmark \\ \hline
\multicolumn{1}{l|}{Panopto}      & \cmark          &                         & \cmark &                         & \cmark \\ \hline
\multicolumn{1}{l|}{Piazza}       & \cmark          &                         &                         & \xmark                         & \cmark \\ \hline
\multicolumn{1}{l|}{Skype for Business}  & \cmark &
  \cmark &
  \cmark &
  \xmark &
  \cmark \\ \hline
\multicolumn{1}{l|}{Slack}       & \cmark           & \xmark                         &                         & \cmark & \cmark \\ \hline
\multicolumn{1}{l|}{Zoom}                   & \xmark                         & \xmark                         & \xmark                         & \xmark                     \\ \hline
\multicolumn{1}{l|}{Zoho}       & \cmark            & \cmark & \xmark                         &                         & \cmark
\end{tabular}
\caption{The breakdown of location data in privacy policies. Active tracking means A-GPS or WiFi location tracking on a continuous basis. Inferred location data includes IP address-based locations and other inferences.}
\label{app:tab:location-privacy}
\end{table*}

\begin{table*}[ht]
    \centering
    \begin{tabular}{cccccc}
    \rowcolor{tableheadcolor} \multicolumn{6}{c}{\textbf{Domains Contacted}} \\ \toprule
        \rowcolor{tableheadcolor} \textbf{BlueJeans} & \textbf{Jitsi}$^\ast$ & \textbf{MS Teams} & \textbf{Slack}  & \textbf{WebEx} & \textbf{Zoom} \\ \midtopline
        bluejeans.com &  & microsoft.com & \textcolor{blue}{\textbf{chime.aws.com}} (Video conferencing)  & webex.com & zoom.us \\
\textcolor{blue}{\textbf{hockeyapp.net}} (Analytics) &  & msedge.com    & \textcolor{blue}{\textbf{gravatar.com}} (Graphic avatars)   &           &         \\
\textcolor{blue}{\textbf{mixpanel.com}} (Analytics)  &  &               & slack.com      &           &         \\
\textcolor{blue}{\textbf{nr-data.net}} (Analytics)   &  &               & slack-edge.com &           &         \\
              &  &               & slack-ims.com  &           &        
        \end{tabular}
        \caption{{The domains contacted by each platform in our network analysis. \textcolor{blue}{\textbf{Blue}} domains are third-party domains. $^\ast$ = Jitsi requires specifying a server domain to connect to, and does not connect to other domains.}}
        \label{tab:network-traffic}
\end{table*}

{
\def\colorModel{hsb} 

\newcommand{\centeredtext}[1]{\begin{tabular}{l} #1 \end{tabular}}

\newcommand\ColCell[1]{
  \pgfmathparse{#1<50?1:0}  
    \ifnum\pgfmathresult=0\relax\color{white}\fi
  \pgfmathsetmacro\compA{0}      
  \pgfmathsetmacro\compB{#1/10} 
  \pgfmathsetmacro\compC{1}      
  \edef\x{\noexpand\centering\noexpand\cellcolor[\colorModel]{\compA,\compB,\compC}}\x #1
  } 
\newcolumntype{E}{>{\collectcell\ColCell}r<{\endcollectcell}}  

\begin{table*}[ht]
\begin{tabular}{l@{}cccccccE}
    \rowcolor{tableheadcolor}
                                          Permission &  BlueJeans &  Jitsi &  Slack &  Teams &  WebEx (M) & WebEx (T) &  Zoom &  \multicolumn{1}{r}{Totals} \\ \midtopline
Version & 41.1813 & 20.2.3 & 206.10 & 1416 & 40.6.1 & 4.11.241 & 5.1.27838 \\
                         access Bluetooth settings &                                  \cmark &             \xmark &            \xmark &                                 \cmark &                       \xmark &                      \xmark &                              \cmark &    3 \\
                           access download manager &                                  \xmark &             \xmark &            \xmark &                                 \cmark &                       \xmark &                      \xmark &                              \xmark &  1 \\
 add/modify calendar events and send email... &                                  \xmark &             \cmark &            \xmark &                                 \xmark &                       \xmark &                      \xmark &                              \cmark &    2 \\
                            add or remove accounts &                                  \cmark &             \xmark &            \xmark &                                 \cmark &                       \cmark &                      \cmark &                              \cmark &    5 \\
              approximate location (network-based) &                                  \cmark &             \xmark &            \xmark &                                 \cmark &                       \cmark &                      \cmark &                              \cmark &    5 \\
                       change network connectivity &                                  \xmark &             \xmark &            \xmark &                                 \cmark &                       \xmark &                      \xmark &                              \xmark &  1 \\
                        change your audio settings &                                  \cmark &             \cmark &            \cmark &                                 \cmark &                       \cmark &                      \cmark &                              \cmark &    7 \\
                 connect and disconnect from Wi-Fi &                                  \xmark &             \xmark &            \xmark &                                 \xmark &                       \xmark &                      \cmark &                              \xmark &  1 \\
                                 control vibration &                                  \cmark &             \xmark &            \cmark &                                 \cmark &                       \xmark &                      \cmark &                              \cmark &    5 \\
                 create accounts and set passwords &                                  \xmark &             \xmark &            \xmark &                                 \cmark &                       \cmark &                      \cmark &                              \xmark &    3 \\
                       directly call phone numbers &                                  \cmark &             \xmark &            \xmark &                                 \cmark &                       \cmark &                      \cmark &                              \cmark &    5 \\
                          disable your screen lock &                                  \cmark &             \xmark &            \xmark &                                 \xmark &                       \xmark &                      \xmark &                              \xmark &  1 \\
               download files without notification &                                  \xmark &             \xmark &            \xmark &                                 \cmark &                       \xmark &                      \xmark &                              \xmark &  1 \\
                              draw over other apps &                                  \cmark &             \cmark &            \xmark &                                 \cmark &                       \cmark &                      \cmark &                              \cmark &    6 \\
                        expand/collapse status bar &                                  \xmark &             \xmark &            \xmark &                                 \cmark &                       \xmark &                      \xmark &                              \xmark &  1 \\
                       find accounts on the device &                                  \cmark &             \xmark &            \cmark &                                 \cmark &                       \cmark &                      \cmark &                              \cmark &    6 \\
                               full network access &                                  \cmark &             \cmark &            \cmark &                                 \cmark &                       \cmark &                      \cmark &                              \cmark &    7 \\
                                 install shortcuts &                                  \xmark &             \xmark &            \xmark &                                 \xmark &                       \cmark &                      \xmark &                              \xmark &  1 \\
 modify or delete USB storage &                                  \cmark &             \cmark &            \cmark &                                 \cmark &                       \cmark &                      \cmark &                              \cmark &    7 \\
                            modify system settings &                                  \xmark &             \xmark &            \xmark &                                 \xmark &                       \xmark &                      \xmark &                              \cmark &  1 \\
                              modify your contacts &                                  \xmark &             \xmark &            \xmark &                                 \cmark &                       \xmark &                      \cmark &                              \xmark &    2 \\
                       pair with Bluetooth devices &                                  \cmark &             \cmark &            \cmark &                                 \cmark &                       \cmark &                      \cmark &                              \cmark &    7 \\
          precise location (GPS \& network-based) &                                  \cmark &             \xmark &            \xmark &                                 \cmark &                       \cmark &                      \xmark &                              \cmark &    4 \\
                      prevent device from sleeping &                                  \cmark &             \cmark &            \cmark &                                 \cmark &                       \cmark &                      \cmark &                              \cmark &    7 \\
 read calendar events plus confidential info... &                                  \cmark &             \cmark &            \xmark &                                 \xmark &                       \cmark &                      \cmark &                              \cmark &    5 \\
                    read phone status and identity &                                  \cmark &             \cmark &            \cmark &                                 \xmark &                       \cmark &                      \cmark &                              \cmark &    6 \\
                                read sync settings &                                  \xmark &             \xmark &            \xmark &                                 \xmark &                       \cmark &                      \xmark &                              \xmark &  1 \\
             read USB storage &                                  \cmark &             \cmark &            \cmark &                                 \cmark &                       \cmark &                      \cmark &                              \cmark &    7 \\
                                read your contacts &                                  \cmark &             \xmark &            \cmark &                                 \cmark &                       \cmark &                      \cmark &                              \cmark &    6 \\
                        receive data from Internet &                                  \cmark &             \cmark &            \cmark &                                 \cmark &                       \cmark &                      \cmark &                              \cmark &    7 \\
                                      record audio &                                  \cmark &             \cmark &            \cmark &                                 \cmark &                       \cmark &                      \cmark &                              \cmark &    7 \\
                              reorder running apps &                                  \xmark &             \xmark &            \xmark &                                 \xmark &                       \cmark &                      \xmark &                              \xmark &  1 \\
                             retrieve running apps &                                  \xmark &             \xmark &            \xmark &                                 \xmark &                       \cmark &                      \xmark &                              \xmark &  1 \\
                                    run at startup &                                  \cmark &             \xmark &            \cmark &                                 \cmark &                       \xmark &                      \xmark &                              \xmark &    3 \\
                             send sticky broadcast &                                  \cmark &             \xmark &            \xmark &                                 \xmark &                       \cmark &                      \xmark &                              \cmark &    3 \\
                          take pictures and videos &                                  \cmark &             \cmark &            \xmark &                                 \cmark &                       \cmark &                      \cmark &                              \cmark &    6 \\
                            toggle sync on and off &                                  \xmark &             \xmark &            \xmark &                                 \xmark &                       \cmark &                      \xmark &                              \xmark &  1 \\
                               uninstall shortcuts &                                  \xmark &             \xmark &            \xmark &                                 \xmark &                       \cmark &                      \xmark &                              \xmark &  1 \\
                        use accounts on the device &                                  \cmark &             \xmark &            \xmark &                                 \cmark &                       \xmark &                      \xmark &                              \cmark &    3 \\
                            view Wi-Fi connections &                                  \cmark &             \cmark &            \xmark &                                 \cmark &                       \cmark &                      \cmark &                              \cmark &    6 \\
                          view network connections &                                  \cmark &             \cmark &            \cmark &                                 \cmark &                       \cmark &                      \cmark &                              \cmark &    7 \\
    \rowcolor{tableheadcolor}
Totals                                                 &       \cellcolor[hsb]{0,0.60,1}26 &    \cellcolor[hsb]{0,0.375,1}15 &    \cellcolor[hsb]{0,0.35,1}14 &    \cellcolor[hsb]{0,0.7,1}28 &    \cellcolor[hsb]{0,0.7,1}28 & \cellcolor[hsb]{0,0.57,1}23 &  \cellcolor[hsb]{0,0.60,1}26 &   \multicolumn{1}{r}{138} \\ 
\end{tabular}
\caption{\textbf{Permissions Requested by Android Apps.} We tabulated the permissions requested by the different applications when installed on Android. We sourced permissions from the app listings on Google Play.}
\label{tab:permissions}
\end{table*}
}

\section{Android Permissions}
\crefalias{section}{appendix}
\label{app:security:permissions}
When an individual uses an app on either iOS (\apple) or Android (\android), the user is promoted to allow or deny various permissions that the app requests. Such permissions vary from the relatively innocuous (take/view photos) to fairly powerful (control system settings). Depending on the permissions model of the operating system the user may be prompted at install time \android, run time \android/\apple, or as additional permissions are requested by the app \android/\apple. Each app's use case requires tailored permissions. However, a developer is free to request permissions that are not essential for the underlying project but instead further the developer's business interests, for example, facilitating the capture of user data for later resale to third parties. Such extra permissions are also inconsistent with the security principle of least privilege to which mobile OS developers attempt to adhere: no app should have more permissions than it requires, to do otherwise increases the attack surface area. Thus, while the presence of any particular permission granted to an app may not on its own be reason for concern, over-broad requests for permissions pose problems for both security and privacy.

\section{Network Traffic Domains}
\crefalias{section}{appendix}
\label{app:security:network}

We present the list of first-party and third-party domains contacted by each desktop platform during our network traffic analysis, along with the type of service provided by each third party, in \cref{tab:network-traffic}. The Jitsi client only contacts the domain that a user specifies, so we do not report any contacted domains for it.


\minihead{Results}
We collected the set of requested permissions for the Android versions of the products from each app's page on the Google Play Store website.
As iOS apps obtain permissions when the associated action is attempted (rather than at install time, as on Android), gathering the equivalent data for the operating system would require exhaustively interacting with all app features, a task we leave to future work.
We tabulate the full set of results in \cref{tab:permissions}.

We found that the applications requested between fourteen and twenty-eight different permissions, with the average application requesting twenty-three permissions.
More or fewer permissions requested is not inherently better or worse.
However, apps with more permissions do carry a higher risk of violating the principle of least privilege, and therefore of facilitating privacy and security violations.

\section{Binary Security Feature Descriptions}
\crefalias{section}{appendix}
\label{app:security:binary}

\minihead{SafeSEH (Windows Only)} Safe Structured Exception Handling (SafeSEH) is a mechanism to ensure that only authorized exception handlers execute~\cite{sotirov2009bypassing}. A binary with SafeSEH contains a list of exception handlers, which the kernel stores in a protected list when the program begins execution. If an exception is thrown, the kernel checks if the handler is pre-approved and, if so, allows the handler to execute.

\minihead{DEP/NX} Data Execution Prevention techniques separate areas of memory that contain data and those that contain code, restricting the user from executing code contained in areas marked for data. No Execute bit (NX) is a CPU implementation of the DEP concept. A binary with support for DEP/NX has appropriately marked memory regions.

\minihead{ASLR} \Gls{aslr} randomizes the layout of a binary when the operating system loads it into a virtual address space~\cite{team2003pax}.
Its efficacy is tied to the number of possible different layouts. 32-bit operating systems traditionally reserved only randomized 8 bits, giving an attacker a 1/256 chance of guessing the layout \cite{aslrng}.
Modern 64-bit versions of MacOS and Windows support randomizing up to 16 and 19 bits of entropy respectively (corresponding to 66k and 524k guesses). Notably, Windows requires two compile time flags to be set to enable 19 bit ASLR, without which at most 14 bits can be randomized (16k guesses) \cite{aslrwindows, defeatingaslr}.

\minihead{CFI} Control flow integrity refers to a class of mitigations that aim to prevent an attacker from redirecting program flow~\cite{abadi2009control}. Microsoft’s implementation of this concept, Control Flow Guard (CFG), adds a check before {\color{blue}\texttt{call}} instructions (that transfer execution to a function) that do not have static arguments. The check cross-references a data structure that stores the start address of all valid functions, and throws an exception if the program execution would otherwise be transferred to an invalid address \cite{exploringcfg}.

\minihead{Code Signing} Code signing provides a chain-of-trust to validate authorship of a binary. Both Microsoft and Apple provide services for third-party developers to sign their applications. Both prevent users of their most modern operating system versions from executing unsigned binaries absent explicit acknowledgment.

\minihead{Stack Canaries} Stack canaries are a compile-time modification that allow a program to detect a subset of buffer overflow attacks~\cite{cowan1998stackguard}. A canary value is placed on the stack between a stack frame’s return pointer and other variables.  Before a program uses the return pointer on the bottom of the stack, it first checks the contents of the stack canary against a known value. If the value has been altered, the stack has been tampered with, and the program will terminate with an error.

\minihead{Architecture Width} While the architecture for which a binary is compiled is not inherently a security feature, 32-bit binaries are unable to use many modern OS security measures. To this end, MacOS 15 (Oct 2019) ceased support for 32-bit binaries. We therefore checked that the Windows software packages were 64-bit. (Note that 32-bit software packages that developers migrate to 64-bit architectures may exhibit pathologies~\cite{s4issues}. Thus, packages that have recently transitioned to 64-bit architectures merit further caution.)

\end{document}